%% file: marfcat-sate4-arxiv.tex
\documentclass{easychair}

\usepackage[figure,linesnumbered,norelsize]{algorithm2e}
\usepackage{fancyvrb}

\usepackage{makeidx}

\makeindex

\input{commands}

\newcommand{\ngram}{$n$-gram}
\newcommand{\marfcat}{MARFCAT\index{MARFCAT}\index{MARF!Applications!MARFCAT}}

\newcommand
	{\cve}
	[1]
	{\href{http://cve.mitre.org/cgi-bin/cvename.cgi?name=#1}{#1}\index{CVE!#1}}

\newcommand
	{\cwe}
	[1]
	{#1\index{CWE!#1}}

\newcommand
	{\chromeCase}
	[1]
	{Chrome 5.0.375.#1\index{Chrome!5.0.375.#1}\index{Test cases!Chrome 5.0.375.#1}}

\newcommand
	{\wiresharkCase}
	[1]
	{Wireshark 1.2.#1\index{Wireshark!1.2.#1}\index{Test cases!Wireshark 1.2.#1}}

\newcommand
	{\dovecotCase}
	[1]
	{Dovecot\index{Dovecot #1}\index{Test cases!Dovecot #1}}

\newcommand
	{\tomcatCase}
	[1]
	{Tomcat 5.5.#1\index{Tomcat!5.5.#1}\index{Test cases!Tomcat 5.5.#1}}

\newcommand
	{\jettyCase}
	[1]
	{Jetty 6.1.#1\index{Jetty!6.1.#1}\index{Test cases!Jetty 6.1.#1}}

\newcommand
	{\pebbleCase}
	{Pebble\index{Pebble}\index{Test cases!Pebble 2.5-M2}}

\newcommand
	{\wordpressCase}
	[1]
	{Wordpress 2.#1\index{Wordpress!2.#1}\index{Test cases!Wordpress 2.#1}}

\newcommand{\NIST}{}
\newcommand{\SATE}{SATE IV}

\begin{document}

\title{MARFCAT: Transitioning to Binary and Larger Data Sets of {\SATE}}

\titlerunning{MARFCAT: A MARF Approach to {\SATE}}

\author{Serguei A. Mokhov$^{1,2}$, Joey Paquet$^{1}$, Mourad Debbabi$^{1}$, Yankui Sun$^{2}$\\
\affiliation{$^{1}$Concordia University}\\
\affiliation{Montreal, QC, Canada}\\
\affiliation{\url{{mokhov,paquet,debbabi}@encs.concordia.ca}}
\and
\affiliation{$^{2}$Tsinghua University}\\
\affiliation{Beijing, China}\\
\affiliation{\url{syk@mail.tsinghua.edu.cn}}
}

\authorrunning{Mokhov, Paquet, Debbabi, Sun}

\maketitle

\NIST{}
\NIST{}

\begin{abstract}
We present a second iteration of a machine learning approach to static
code analysis and fingerprinting for weaknesses related to security, software engineering,
and others using the open-source {\marf} framework and the {\marfcat} application
based on it for the NIST's {\SATE} static analysis tool exposition workshop's data
sets that include additional test cases, including new large synthetic cases.
To aid detection of weak or vulnerable code, including source or binary on different
platforms the
machine learning
approach proved to be fast and accurate to
for such tasks where other tools are either much slower or have much smaller recall
of known vulnerabilities.
We use signal and NLP processing techniques in our approach to accomplish
the identification and classification tasks.
{\marfcat}'s design from the beginning in 2010 made is independent of the
language being analyzed, source code, bytecode, or binary. In this follow up
work with explore some preliminary results in this area.
We evaluated also additional algorithms that were used to process the data.
\end{abstract}

\tableofcontents
\listoftables
\listofalgorithms

\section{Introduction}
\label{sect:introduction}

This is a follow up work on the first incarnation of {\marfcat}
detailed in
\cite{marfcat-arxiv,marfcat-sate2010-nist}.
Thus, the majority of the results content here addresses the newer iteration
duplicating only the necessary background and methodology information (reduced).
The reader is deferred to consult the expanded background
information and results in that previous work freely accessible
online (and the arXiv version of that is still occasionally updated).

\nocite{marf}
We elaborate on the details of the expanded methodology
and the corresponding results of application of the machine
learning techniques along with signal and NLP processing
to static source and binary code analysis in search for weaknesses and
vulnerabilities.
We use the tool, named {\em {\marfcat}}, a {\marf}-based Code Analysis Tool
\cite{marfcat-app},
first exhibited at the Static Analysis Tool Exposition (SATE) workshop in 2010
\cite{nist-samate-sate2010}
to machine-learn from the (Common Vulnerabilities and Exposures) CVE-based vulnerable as well
as synthetic CWE-based cases to verify the fixed versions as well as
non-CVE based cases from the projects written in same programming languages.
The $2^{nd}$ iteration of this work was prepared
based on {\SATE} \cite{nist-samate-sate4} and uses its updated data set and application.
On the NLP side,
we employ simple classical NLP techniques ($n$-grams and various smoothing
algorithms), also combined with machine learning for
novel non-NLP applications of detection, classification, and reporting of
weaknesses related to vulnerabilities or bad coding practices found in artificial
constrained languages, such as programming languages and their compiled counterparts. 
We compare and contrast the NLP approach to the signal processing approach
in our results summary and illustrate concrete results and for the same test cases.

We claim that the presented machine learning approach is novel and highly beneficial in static analysis
and routine testing of any kind of code, including source code and binary
deployments for its efficiency in terms of speed, relatively high precision, robustness,
and being a complimentary tool to other approaches that do in-depth semantic
analysis, etc. by prioritizing those tools' targets. All that can be used in automatic
manner in distributed and scalable diverse environments to ensure the
code safety, especially the mission critical software code in all kinds
of systems. It uses spectral, acoustic and language models to learn
and classify such a code.

This document, like its predecessor, is a ``rolling draft'' with several updates expected
to be made as the project progresses beyond {\SATE}. It is accompanied
with the updates to the open-source {\marfcat} tool itself \cite{marfcat-app}.

\subsection*{Organization}

The related work, some of the present methodology is based on, is referenced in \xs{sect:related-work}.
The data sets are described in \xs{sect:data-sets}.
The methodology summary is in \xs{sect:methodology}.
We present some of the results
in \xs{sect:results} from the SAMATE reference test data set.
Then we present a brief summary, description of the limitations of the current realization of the approach and
concluding remarks in \xs{sect:conclusion}.
In the Appendix there are classification result tables for specific test cases
illustrating top results by precision.

\section{Related Work}
\label{sect:related-work}

To our knowledge this was the first time a machine learning approach
was attempted to static code analysis with the first results demonstrated
during the SATE2010 workshop
\cite{marfcat-arxiv,marfcat-app,nist-samate-sate2010}.
In the same year, a somewhat similar approach independently was presented
\cite{learn-class-vulns-predict-exploits-2010} for vulnerability classification
and prediction using machine learning and SVMs, but working with a different
set of data.

Additional
related work (to various degree of relevance or use)
is further listed (this list is not exhaustive).
A taxonomy of Linux kernel vulnerability solutions in terms of
patches and source code as well as categories for both are found in
\cite{linux-vuln-sols-cisse07}.
The core ideas and principles behind the {\marf}'s pipeline and testing
methodology for various algorithms in the pipeline adapted to this case
are found in \cite{marf-c3s2e08,marf-nlp-framework} as it was the
easiest implementation available to accomplish the task.
There also one can find the majority of the core options
used to set the configuration for the pipeline in terms of algorithms used.
A binary analysis using machine learning approach for quick scans for
files of known types in a large collection of files is described in
\cite{marf-file-type}.
This includes the NLP and machine learning for NLP tasks in
DEFT2010 \cite{marf-deft,marf-deft-complete-results} with the
corresponding \api{DEFT2010App} and its predecessor for hand-written image processing
\api{WriterIdentApp} \cite{marf-writer-ident}.
Tlili's 2009 PhD thesis covers topics on automatic detection of safety and security
vulnerabilities in open source software \cite{tlili-phd-vuln-detect-oss-2009}.
Statistical analysis, ranking, approximation, dealing with uncertainty,
and specification inference in static code analysis are found in the
works of Engler's team \cite{stats-spec-within,correlation-error-ranking,z-ranking-2003}.
Kong et al. further advance static analysis (using parsing, etc.) and specifications
to eliminate human specification from the static code analysis in 
\cite{no-human-spec-static-analysis}.
Spectral techniques are used for pattern scanning in malware detection by Eto et al. in
\cite{malware-spectrum-analysis-2009}.
Some researchers propose a general data mining system for incident analysis with data mining
engines in \cite{incident-analysis-nicter-data-mining-wistdcs-2008}.
Hanna et al. describe a synergy between static and dynamic analysis for the detection
of software security vulnerabilities in \cite{synergy-static-dynamic-2009} paving the way
to unify the two analysis methods.
Other researchers propose a MEDUSA system for metamorphic malware dynamic analysis
using API signatures in \cite{medusa-malanal-api-sin-2010}.
Some of the statistical NLP techniques we used, are described at length in \cite{manning2002}.
BitBlaze (and its web counterpart, WebBlaze) are other recent types of tools that
to static and dynamic binary code analysis for vulnerabilities fast, developed
at Berkeley \cite{bitblaze,webblaze}.
For wavelets, for example, Li et al. \cite{net-app-id-wavlets-k-means-icis2009}
have shown wavelet transforms and
$k$-means classification can be used to identify communicating applications
on a network fast and is relevant to our study of the code in any form,
text or binary.

\section{Data Sets}
\label{sect:data-sets}

We use the SAMATE data set to practically validate our approach.
The SAMATE reference data set contains {\C}/{\cpp}, {\java}, and {\php} language
tracks comprising CVE-selected cases as well as stand-alone cases
and the large generated synthetic {\C} and {\java} test cases (CWE-based, with
a lot of variants of different known weaknesses). {\SATE} expanded some cases
from SATE2010 by increasing the version number, and dropped some other cases (e.g., Chrome).

The {\C}/{\cpp} and {\java} test cases of various client and server OSS software
are compilable into the binary and object code, while the synthetic
{\C} and {\java} cases generated for various CWE entries provided for
greater scalability testing (also compilable).
The CVE-selected cases had a vulnerable version of a software in question
with a list of CVEs attached to it, as well as the most known fixed version
within the minor revision number. One of the goals for the CVE-based cases
is to detect the known weaknesses outlined in CVEs using static code analysis
and also to verify if they were really fixed in the ``fixed version'' \cite{nist-samate-sate4}.
The cases with known CVEs and CWEs were used as the training models
described in the methodology. The summary below is a union of the data
sets from SATE2010 and {\SATE}.
The preliminary list of the CVEs that the organizers expect to locate
in the test cases were collected from the NVD \cite{nist,nist-samate-sate4} for
\href
	{http://web.nvd.nist.gov/view/vuln/search-results?adv_search=true&cves=on&cpe_product=cpe:/a:wireshark:wireshark&cpe_version=cpe:/a:wireshark:wireshark:1.2.0}
	{\wiresharkCase{0}},
\href
	{http://web.nvd.nist.gov/view/vuln/search-results?adv_search=true&cves=on&cpe_product=cpe%3A%2Fa%3Adovecot%3Adovecot&cpe_version=cpe%3A%2Fa%3Adovecot%3Adovecot%3A1.2.0}
	{\dovecotCase{ 1.2.0}},
\href
	{http://web.nvd.nist.gov/view/vuln/search-results?adv_search=true&cves=on&cpe_product=cpe%3A%2Fa%3Aapache%3Atomcat&cpe_version=cpe%3A%2Fa%3Aapache%3Atomcat%3A5.5.13}
	{\tomcatCase{13}},
\href
	{http://web.nvd.nist.gov/view/vuln/search-results?adv_search=true&cves=on&cpe_product=cpe%3A%2Fa%3Amortbay%3Ajetty&cpe_version=cpe%3A%2Fa%3Amortbay%3Ajetty%3A6.1.16}
	{\jettyCase{16}}, and
\href
	{http://web.nvd.nist.gov/view/vuln/search-results?adv_search=true&cves=on&cpe_product=cpe%3A%2Fa%3Awordpress%3Awordpress&cpe_version=cpe%3A%2Fa%3Awordpress%3Awordpress%3A2.0}
	{\wordpressCase{0}}.
The specific test cases with versions and language at the time included CVE-selected:

\begin{itemize}
	\item {\C}: \wiresharkCase{0} (vulnerable) and \wiresharkCase{18} (fixed, up from \wiresharkCase{9} in SATE2010)
	\item {\C}: \dovecotCase{ 1.2.0} (vulnerable) and \dovecotCase{ 1.2.17} (fixed)
	\item {\cpp}: \chromeCase{54} (vulnerable) and \chromeCase{70} (fixed)
	\item {\java}: \tomcatCase{13} (vulnerable) and \tomcatCase{33} (fixed, up from \tomcatCase{29} in SATE2010)
	\item {\java}: \jettyCase{16} (vulnerable) and \jettyCase{26} (fixed)
	\item {\php}: \wordpressCase{0} (vulnerable) and \wordpressCase{2.3} (fixed)
\end{itemize}

\noindent
originally non-CVE selected in SATE2010:

\begin{itemize}
	\item {\C}: \dovecotCase{2.0.beta6.20100626}
	\item {\java}: {\pebbleCase} 2.5-M2
\end{itemize}

\noindent
Synthetic CWE cases produced by the SAMATE team:

\begin{itemize}
	\item {\C}: Synthetic {\C} covering 118 CWEs and $\approx 60$K files
	\item {\java}: Synthetic {\java} covering $\approx 50$ CWEs and $\approx 20$K files
\end{itemize}

\section{Methodology}
\label{sect:methodology}

In this section we outline the methodology of our approach to
static source code analysis. Most of this methodology is an
updated description from
\cite{marfcat-arxiv}.
The line number determination methodology is also detailed in
\cite{marfcat-arxiv,nist-samate-sate2010},
but is not replicated here.
Thus,
the methodology's principles overview is described in
\xs{sect:core-principles}, the knowledge base construction is in \xs{sect:kb},
machine learning categories in \xs{sect:ml-cats}, and
the high-level algorithmic description is in \xs{sect:basic-methodology}.

\subsection{Methodology Overview}
\label{sect:core-principles}

The core methodology principles include:

\begin{itemize}
\item Machine learning and dynamic programming
\item Spectral and signal processing techniques
\item NLP {\ngram} and smoothing techniques (add-$\delta$, Witten-Bell, MLE, etc.)
\end{itemize}

We use signal processing techniques, i.e. presently we do not parse
or otherwise work at the syntax and semantics levels. We treat the source code as a ``signal'',
equivalent to binary, where each {\ngram} ($n=2$ presently, i.e. two  
consecutive characters or, more generally, bytes) are used to  
construct a sample amplitude value in the signal. In the NLP pipeline,
we similarly treat the source code as a ``characters'',
where each {\ngram} ($n=1..3$) is used to  
construct the language model.

We show the system the examples of files with weaknesses and {\marfcat} learns them
by computing spectral signatures using signal processing techniques
or various language models (based on options)
from CVE-selected test cases.
When some of the mentioned techniques are applied (e.g., filters, silence/noise removal, other
preprocessing and feature extraction techniques), the line number information
is lost as a part of this process.

When we test, we compute either
how similar or distant each file is from the known trained-on weakness-laden files
or compare trained language models with the unseen language fragments in the NLP pipeline.
In part, the methodology can approximately be seen as some fuzzy signature-based ``antivirus'' or IDS software systems
detect bad signature, except that with a large number of machine learning and signal processing
algorithms and fuzzy matching, we test to find out which combination gives the highest
precision and best run-time.

At the present, however, we are looking at the whole files
instead of parsing the finer-grain details of patches and weak code fragments.
This aspect lowers the precision, but is relatively fast to scan all the 
code files.

\subsection{CVEs and CWEs -- the Knowledge Base}
\label{sect:kb}

The CVE-selected test cases serve as a source of the knowledge
base to gather information of how known weak code ``looks like''
in the signal form
\cite{marfcat-arxiv},
which we store as spectral signatures
clustered per CVE or CWE (Common Weakness Enumeration).
The introduction by the SAMATE team of a large synthetic code base with CWEs,
serves as a part of knowledge base learning as well.
Thus, we:

\begin{itemize}
\item Teach the system from the CVE-based cases 
\item Test on the CVE-based cases
\item Test on the non-CVE-based cases
\end{itemize}

\noindent
For synthetic cases we do similarly:

\begin{itemize}
\item Teach the system from the CWE-based synthetic cases 
\item Test on the CWE-based synthetic cases
\item Test on the CVE and non-CVE-based cases for CWEs from synthetic cases
\end{itemize}

We create index files in XML in the format similar to that of SATE to index
all the file of the test case under study. The CVE-based cases after the
initial index generation are manually annotated from the NVD database
before being fed to the system.
The script that does the initial index
gathering in the OSS distribution of {\marfcat} is called \file{collect-files-meta.pl}
written in {\perl}. The synthetic cases required a special modification to
that resulting in \file{collect-files-meta-synthetic.pl} where there are no
CVEs to fill in but CWEs alone, with the auto-prefilled explanations since
the information in the synthetic cases is not arbitrary and controlled for
identification.

\subsection{Categories for Machine Learning}
\label{sect:ml-cats}

The tow primary groups of classes we train and test on
include are naturally the CVEs \cite{nist,niststats}
and CWEs \cite{mitre-cwes}.
The advantages of CVEs is the precision and the associated
meta knowledge from \cite{nist,niststats} can be all aggregated
and used to scan successive versions of the the same software
or derived products (e.g., WebKit in multiple browsers). CVEs are also generally uniquely mapped
to CWEs.
The CWEs as a primary class, however, offer broader categories,
of kinds of weaknesses there may be, but are not yet well
assigned and associated with CVEs, so we observe the loss of
precision.
Since we do not parse, we generally cannot deduce weakness types
or even simple-looking aspects like line numbers where the weak
code may be. So we resort to the secondary categories, that are
usually tied into the first two, which we also machine-learn
along, such as issue types ({\em sink}, {\em path}, {\em fix})
and line numbers.

\subsection{Algorithms}
\label{sect:basic-methodology}

In our methodology we systematically test and select the best (a tradeoff between
speed and accuracy) combination(s) of the algorithm implementations available to
us and then use only those for subsequent testing.
This methodology is augmented with the cases when the knowledge base for the
same code type is learned from multiple sources (e.g., several independent {\C} test cases).

\subsubsection{Signal Pipeline}

Algorithmically-speaking, the steps that are performed in
the machine-learning signal based analysis are in \xf{algo:marfcat-overall-algo}.
The specific algorithms come from the classical literature and other sources and are
detailed in \cite{marf-c3s2e08} and the related works. To be more
specific for this work, the loading typically refers to the interpretation
of the files being scanned in terms of bytes forming amplitude values
in a signal (as an example, 8kHz or 16kHz frequency) using either uni-gram,
bi-gram, or tri-gram approach.
Then, the
preprocessing allows to be none at all (``raw'', or the fastest), normalization,
traditional frequency domain filters, wavelet-based filters, etc.
Feature extraction involves reducing an arbitrary length signal
to a fixed length feature vector of what thought to be the most relevant
features are in the signal (e.g., spectral features in FFT, LPC), min-max
amplitudes, etc. Classification stage is then separated either to train
by learning the incoming feature vectors (usually $k$-means clusters, median clusters, or
plain feature vector collection, combined with e.g., neural network training)
or testing them against the previously learned models.

\begin{algorithm}[hptb]
\SetAlgoLined
\tcp{Construct an index mapping CVEs to files and locations within files}
Compile meta-XML index files from the CVE reports (line numbers, CVE, CWE, fragment size, etc.).
Partly done by a Perl script and partly annotated manually\;
\ForEach{source code base, binary code base}
{
	\tcp{Presently in these experiments we use simple mean clusters of feature vectors
	or unigram language models per default {\marf} specification (\cite{marf-c3s2e08})}
	Train the system based on the meta index files to build the knowledge base (learn)\;
	\Begin
	{
		Load (interpret as a wave signal or $n-gram$)\;
		Preprocess (none, FFT-filters, wavelets, normalization, etc.)\;
		Extract features (FFT, LPC, min-max, etc.)\;
		Train (Similarity, Distance, Neural Network, etc.)\;
	}

	Test on the training data for the same case (e.g., \tomcatCase{13} on \tomcatCase{13}) with the same annotations
	to make sure the results make sense by being high and deduce the best algorithm combinations for the task\;
	\Begin
	{
		Load (same)\;
		Preprocess (same)\;
		Extract features (same)\;
		Classify (compare to the trained $k$-means, or medians, or language models)\;
		Report\;
	}

	Similarly test on the testing data for the same case (e.g., \tomcatCase{13} on \tomcatCase{13}) without the annotations
	as a sanity check\;

	Test on the testing data for the fixed case of the same software (e.g., \tomcatCase{13} on \tomcatCase{33})\;

	Test on the testing data for the general non-CVE case (e.g., \tomcatCase{13} on {\pebbleCase} or synthetic)\;
}
\caption{Machine-learning-based static code analysis testing algorithm using the signal pipeline}
\label{algo:marfcat-overall-algo}
\end{algorithm}

\subsubsection{NLP Pipeline}

The steps that are performed in NLP and
the machine-learning based analysis are presented in \xf{algo:marfcat-overall-nlp-algo}.
The specific algorithms again come from the classical literature (e.g., \cite{manning2002}) and are
detailed in
\cite{marf-nlp-framework} and
the related works. To be more
specific for this work, the loading typically refers to the interpretation
of the files being scanned in terms of {\ngram}s:
uni-gram,
bi-gram, or tri-gram approach and the associated statistical smoothing
algorithms, the results of which (a vector, 2D or 3D matrix) are stored.

\begin{algorithm}[hptb]
\SetAlgoLined
Compile meta-XML index files from the CVE reports (line numbers, CVE, CWE, fragment size, etc.).
Partly done by a Perl script and partly annotated manually\;

\ForEach{source code base, binary code base}
{
	\tcp{Presently in these experiments we use simple 
	 unigram language models per default {\marf} specification (\cite{marf-nlp-framework})}
	Train the system based on the meta index files to build the knowledge base (learn)\;
	\Begin
	{
		Load ({\ngram})\;
		Train (statistical smoothing estimators)\;
	}

	Test on the training data for the same case (e.g., \tomcatCase{13} on \tomcatCase{13}) with the same annotations
	to make sure the results make sense by being high and deduce the best algorithm combinations for the task\;
	\Begin
	{
		Load (same)\;
		Classify (compare to the trained language models)\;
		Report\;
	}

	Similarly test on the testing data for the same case (e.g., \tomcatCase{13} on \tomcatCase{13}) without the annotations
	as a sanity check\;

	Test on the testing data for the fixed case of the same software (e.g., \tomcatCase{13} on \tomcatCase{33})\;

	Test on the testing data for the general non-CVE case (e.g., \tomcatCase{13} on {\pebbleCase} or synthetic)\;
}
\caption{Machine-learning-based static code analysis testing algorithm using the NLP pipeline}
\label{algo:marfcat-overall-nlp-algo}
\end{algorithm}

\subsection{Binary and Bytecode Analysis}

In this iteration we also perform preliminary {\java} bytecode and compiled {\C} code
static analysis and produce results using the same signal processing,
NLP, combined with machine learning 
and data mining techniques.
At this writing, the NIST SAMATE synthetic reference data set for {\java}
and {\C} was used. The algorithms presented in \xs{sect:basic-methodology}
are used as-is in this scenario with the modifications to the index files.
The modifications include removal of the line numbers, source code fragments,
and lines-of-text counts (which are largely meaningless and ignored. The byte
counts may be recomputed and capturing a byte offset instead of a line number
was projected. The filenames of the index files were updated to include
\texttt{-bin} in them to differentiate from the original index files describing
the source code. Another point is at the moment the simplifying assumption is
that each compilable source file {e.g., \texttt{.java} or \texttt{.c}} produce
the corresponding \texttt{.class} and \texttt{.o} files that we examine.
We do not examine inner classes or linked executables or libraries at this
point.

\subsection{Wavelets}

As a part of a collaboration project with Dr. Yankui Sun from Tsinghua University,
wavelet-based signal processing
for the purposes of noise filtering is being introduced with this work
to compare it to no-filtering, or FFT-based classical filtering.
It's been also shown in \cite{net-app-id-wavlets-k-means-icis2009}
that wavelet-aided filtering could be used as a fast preprocessing
method for a network application identification and
traffic analysis \cite{wavelet-traffic-time-series-analysis-iccee-2008}.

We rely in part on the the algorithm and methodology found in 
\cite{near-symm-ortho-wavelet-bases-2001,waveletsoftware-matlab,%
tex-img-retrieval-rcwf-2005,rotation-invar-tex-img-retrieval-rcwf-2006},
and at this point only a separating 1D discrete wavelet transform (SDWT) has been
tested (see \xs{sect:results-wavelets}).

Since the original wavelet implementation \cite{waveletsoftware-matlab} is
in MATLAB
\cite{matlab},
\cite{scholarpedia-matlab},
we used in part the \tool{codegen} tool
from the MATLAB Coder toolbox
\cite{matlab-toolbox-coder},
\cite{matlab-toolbox-coder-codegen}
to generate a rough {\C}/{\cpp} equivalent in order to (manually) translate
some fragments into {\java} (the language of {\marf} and {\marfcat}).
The specific function for up/down sampling used by the wavelets function in
\cite{upfirdn-motorla-cpp} written also {\C}/{\cpp} was translated to {\java}
in {\marf} as well with unit tests added.

\subsection{Demand-Driven Distributed Evaluation with {\gipsy}}

To enhance the scalability of the approach, we convert the {\marfcat} stand-alone
application to a distributed one using an eductive model of computation (demand-driven)
implemented in the General Intensional Programming System ({\gipsy})'s
multi-tier run-time system~\cite{bin-han-10,ji-yi-mcthesis-2011,vassev-mscthesis-05,gipsy-multi-tier-secasa09},
which can be executed distributively using {\jini} (Apache River),
or {\jms}~\cite{unifying-refactoring-jini-jms-dms}.

To adapt the application to the {\gipsy}'s multi-tier architecture, we create a problem-specific
generator and worker tiers (PS-DGT and PS-DWT respectively) for the {\marfcat} application.
The generator(s) produce demands of what needs to be computed in the form of a file (source
code file or a compiled binary) to be evaluated and deposit such demands into a store
managed by the demand store tier (DST) as pending. Workers pickup pending demands from
the store, and them process then (all tiers run on multiple nodes) using a traditional
{\marfcat} instance. Once the result (a \api{Warning} instance) is computed, the PS-DWT deposit it back into
the store with the status set to {\em computed}. The generator ``harvests'' all 
computed results (warnings) and produces the final report for a test cases.
Multiple test cases can be evaluated simultaneously or a single case can be
evaluated distributively. This approach helps to cope with large amounts of data
and avoid recomputing warnings that have already been computed and cached in the
DST.

The initial basic experiment assumes the PS-DWTs have the training sets
data and the test cases available to them from the start (either by a copy or
via an NFS/CIFS-mounted volumes); thus, the distributed evaluation only
concerns with the classification task only as of this version. The follow up
work will remove this limitation.

In this setup a demand represents a file (a path) to scan (actually
an instance of the \api{FileItem} object), which is deposited
into the DST. The PS-DWT picks up that and checks the file per training set
that's already there and returns a \api{ResultSet} object back into the
DST under the same demand signature that was used to deposit the path to scan.
The result set is sorted from the most likely to the list likely with a value
corresponding to the distance or similarity. The PS-DGT picks up the result
sets and does the final output aggregation and saves report in one of the
desired report formats (see \xs{sect:report-formats} picking up the top
two results from the result set and testing against a threshold to accept
or reject the file (path) as vulnerable or not. This effectively splits
the monolithic {\marfcat} application in two halves in distributing the
work to do where the classification half is arbitrary parallel.

Simplifying assumptions:

\begin{itemize}
	\item
	Test case data and training sets are present on each node (physical or
	virtual) in advance (via a copy or a CIFS or NFS volume), so no
	demand driven training occurs, only classification
	\item
	The demand assumes to contain only file information to be  examined
	(\api{FileItem})
	\item
	PS-DWT assumes a single pre-defined configuration, i.e. configuration
	for {\marfcat}'s option is not a part of the demand
	\item
	PS-DWT assume CVE or CWE testing based on its local settings and not
	via the configuration in a demand  
\end{itemize}

\subsection{Export}
\label{sect:report-formats}

\subsubsection{SATE}

By default {\marfcat} produces the report data in the SATE XML format, according
to the {\SATE} requirements. In this iteration other formats are being considered
and realized. To enable multiple format output, the {\marfcat} report generation
data structures were adapted case-based output.

\subsubsection{{\flucid}}

The first one, is {\flucid}, the author Mokhov's PhD topic, a language
to specify and evaluate digital forensic cases by uniformly encoding the evidence
and witness accounts (evidential statement or knowledge base) of any case from
multiple sources (system specs, logs, human accounts, etc.) as a description of
an incident to further perform investigation and event reconstruction. 
Following the
data export in {\flucid} in the preceding
work~\cite{flucid-imf08,flucid-raid2010,marf-into-flucid-cisse08}
we use it asa format for evidential processing of the results produced by {\marfcat}.
The work \cite{flucid-imf08} provides details of the language; it will suffice
to mention here that the report generated by {\marfcat} in {\flucid} is a collection
of warnings as observations with the hierarchical notion of nested context of warning
and location information. These will form an evidential statement in {\flucid}.
The example scenario where such evidence compiled via a {\marfcat} {\flucid} report
would be in web-based applications and web browser-based incident investigations
of fraud, XSS, buffer overflows, etc. linking CVE/CWE-based evidence analysis
of the code (binary or source) security bugs with the associated web-based malware
propagation or attacks to provide possible events where specific attacks can be
traced back to the specific security vulnerabilities.

\subsubsection{SAFES}

The third format, for which the export functionality is not done as of this writing,
SAFES, is the 3rd format for output of the {\marfcat}. SAFES is becoming a standard
to reporting such information and the SATE organizers began endorsing it as an
alternative during {\SATE}.

\subsection{Experiments}

The below is the current summary of the conducted experiments:

\begin{itemize}

\item
Re-testing of the newer fixed versions such as
\wiresharkCase{18} and \tomcatCase{33}.

\item
Half-based testing of the previous versions by reducing the
training set by half and but testing for all known CVEs or
CWEs for \wiresharkCase{18}, \tomcatCase{33}, and
\chromeCase{54}.

\item
Testing the new test cases of \dovecotCase{ 1.2.x},
\jettyCase{x}, and \wordpressCase{x} as well as 
Synthetic {\C} and Synthetic {\java}.

\item
Binary test on the Synthetic {\C} and Synthetic {\java}
test cases.

\item
Performing tests using wavelets for preprocessing.

\end{itemize}

\section{Results}
\label{sect:results}

The preliminary results of application of our methodology are outlined
in this section. We summarize the top precisions per test case using
either signal-processing or NLP-processing of the CVE-based and synthetic cases and their
application to the general cases. Subsequent sections detail some of the
findings and issues of {\marfcat}'s result releases with different versions.
Some experiments we compare the results with the previously obtained ones
\cite{marfcat-arxiv}
where compatible and appropriate.

The results currently are being gradually released in the iterative manner
that were obtained through the corresponding versions of {\marfcat}
as it was being designed and developed.

\subsection{Preliminary Results Summary}
\label{sect:prelim-results-summary}

The results summarize the half-training-full-testing data
vs. that of regular ones reported in
\cite{marfcat-arxiv}.

\begin{itemize}

\item Wireshark:
	\begin{itemize}
	\item
	CVEs (signal): 92.68\%,
	CWEs (signal): 86.11\%,
	\item
	CVEs (NLP): 83.33\%,
	CWEs (NLP): 58.33\%
	\end{itemize}

\item Tomcat:
	\begin{itemize}
	\item
	CVEs (signal): 83.72\%,
	CWEs (signal): 81.82\%,
	\item
	CVEs (NLP): 87.88\%,
	CWEs (NLP): 39.39\%
	\end{itemize}

\item Chrome:
	\begin{itemize}
	\item
	CVEs (signal): 90.91\%,
	CWEs (signal): 100.00\%,
	\item
	CVEs (NLP): 100.00\%,
	CWEs (NLP): 88.89\%
	\end{itemize}

\item Dovecot (new, 2.x):
	\begin{itemize}
	\item
	14 warnings; but it appears all quality or false positive
	\item
	(very hard to follow the code, severely undocumented)
	\end{itemize}

\item Pebble:
	\begin{itemize}
	\item
	none found during quick testing
	\end{itemize}

\item Dovecot 1.2.x:
	(ongoing of this writing)

\item Jetty:
	(ongoing of this writing)

\item Wordpress:
	(ongoing of this writing)

\end{itemize}

What follows are some select statistical measurements of the
precision in recognizing CVEs and CWEs under different configurations
using the signal processing and NLP processing techniques.

``Second guess'' statistics provided to see if the hypothesis that
if our first estimate of a CVE/CWE is incorrect, the next one in line
is probably the correct one. Both are counted if the first guess is
correct.

A sample signal visiusalization in the middle of a vulnerable file
\file{packet-afs.c} in \wiresharkCase{0} to \cve{CVE-2009-2562}
is in \xf{fig:wave-graph-wireshark-1-2-0-packet-afs-c-107} in the wave form.
The low ``dips'' represent the text line endings (coupled with a preceding
character (bytes) in bigrams (two PCM-signed bytes assumed encoded in 8kHz
representing the amplitude; normalized), which are often either semicolons, closing or opening
braces, brackets or parentheses). Only a small fragment is shown of roughly
300 bytes in length to be visually comprehensive of a nature of a signal we
are dealing with.

In \xf{fig:spectrograms-wireshark-1-2-x-packet-afs-c}, there are 3 spectrograms
generated for the same file \file{packet-afs.c}. The first two columns represent
the \cve{CVE-2009-2562}-vulnerable file, both versions are the same with ehanced
contrast to see the detail. The subsequent pairs are of the same file in 
\wiresharkCase{9} and \wiresharkCase{18}, where \cve{CVE-2009-2562} is no longer
present. Small changes are noticeable primarily in the bottom left and top right corners
of the images, and even smaller elsewhere in the images.

\begin{figure}[htpb]
	\begin{center}
	\includegraphics[width=\columnwidth]{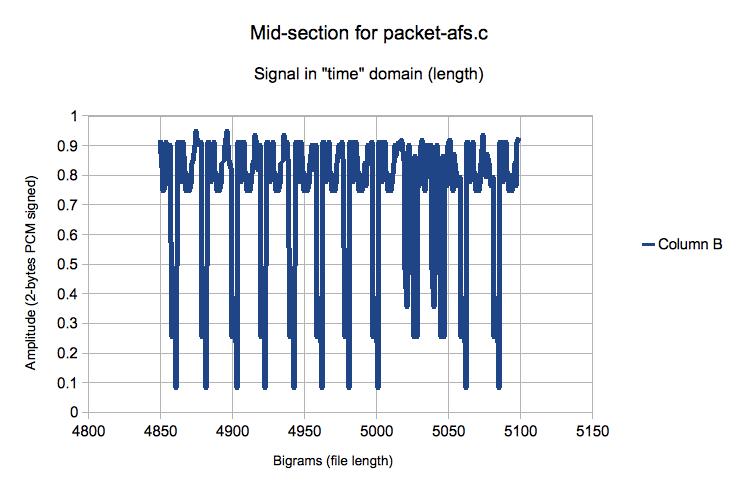}
	\caption{A wave graph of a fraction of the \cve{CVE-2009-2562}-vulnerable \file{packet-afs.c} in \wiresharkCase{0}}
	\label{fig:wave-graph-wireshark-1-2-0-packet-afs-c-107}
	\end{center}
\end{figure}

\begin{figure}[htpb]
	\begin{center}
	\setlength\fboxsep{0pt}
	\setlength\fboxrule{0.5pt}
	\fbox{\includegraphics[angle=90]{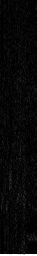}}
	\fbox{\includegraphics[angle=90]{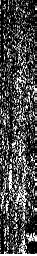}}
	\fbox{\includegraphics[angle=90]{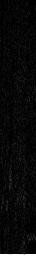}}
	\fbox{\includegraphics[angle=90]{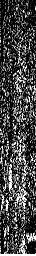}}
	\fbox{\includegraphics[angle=90]{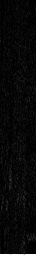}}
	\fbox{\includegraphics[angle=90]{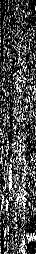}}
	\caption{Spectrograms of \cve{CVE-2009-2562}-vulnerable \file{packet-afs.c} in \wiresharkCase{0}, fixed \wiresharkCase{9} and \wiresharkCase{18}}
	\label{fig:spectrograms-wireshark-1-2-x-packet-afs-c}
	\end{center}
\end{figure}

\subsection{Version SATE-IV.1}

\subsubsection{Half-Training Data For Training and Full For Testing}

This is one of the experiment per discussion with Aurelien Delaitre and SATE organizers.
The main idea is to test robustness and precision of the MARFCAT
approach by artificially reducing known weaknesses (their locations)
to learn from by 50\%, but test on the whole 100\% to see how
much does precision degrade with such a reduction.
Specifically,
we supply only CWE classes testing for this experiment (CVE classes make little sense).
Only the first 50\% of the entries entries were used for training
for \wiresharkCase{0}, \tomcatCase{13}, and \chromeCase{54}, while the full 100\%
were used to test the precision changes. The below are the results.

It should be noted that CWE classification is generally less accurate
due to a lot of dissimilar items ``stuffed'' (by NVD) into very broad categories
such as NVD-CWE-Other and NVD-CWE-noinfo when the data were collected. Additionally, since we arbitrarily picked
the first 50\% of the training data, some of the CWEs simply were left out completely
and not trained on if they were entirely in the omitted half, so their
individual precision is obviously 0\% when tested for.

The archive contains the .log and the .xml files (the latter for now
are in SATE format only with the scientific notation +E3 removed).
The best reports are:

\file{report-cweidnoprepreprawfftcheb-wireshark-1.2.0-half-train-cwe.xml}

\file{report-cweidnoprepreprawfftdiff-wireshark-1.2.0-half-train-cwe.xml}

The experiments are subdivided into regular (signal) and NLP based
testing.

\paragraph{Signal}

\begin{itemize}
\item
\wiresharkCase{0}:

Reduction of the training data by half resulted in $\approx14\%$ precision
drop compared to the previous result (best 86.11\% see the NIST
report \cite{marfcat-sate2010-nist}, vs. 72.22\% overall).

New results (by algorithms, then by CWEs):

\input{results/sate4/wireshark120-cwe-half-training-full-testing}
\normalsize

\item
\tomcatCase{13}:

Drop from 81.82\% (see NIST report's Table 7, p. 70) to 75\% top result as a result
(about 7 points) of training data reduction by 50\%.

New precision estimates:

\input{results/sate4/tomcat13-cwe-half-train-full-test}
\normalsize

\item
\chromeCase{54}:

Chrome result is for completeness even though it is not a test case for {\SATE}.
Chrome is poor for some reason---drop from 100\% (Table 5, p. 68) to 44.44\%, but it's
only 9 entries. The first result in the table is erroenous, i.e., with a poor recall (the sum of $2+0<9$,
should be total 9).

\input{results/sate4/chrome54-cwe-half-train-full-test}
\normalsize

\end{itemize}

\paragraph{NLP}

Generally this genre of classification was poor as before in this
experiment, all around 40-45\% percent precision.

\begin{itemize}

\item
\wiresharkCase{0}:

New results (by algorithms, then by CWEs):

\input{results/sate4/wireshark120-cwe-half-training-full-testing-nlp}
\normalsize

\item
\tomcatCase{13}:

Intriguingly, the best result is higher than with all of the date
in the past report (42.42\% below vs. previous 39.39\%).

\input{results/sate4/tomcat13-cwe-half-train-full-test-nlp}
\normalsize

\item
\chromeCase{54}:

Here drop is twice as much ($\approx44\%$ vs. 88\%).

\input{results/sate4/chrome54-cwe-half-train-full-test-nlp}
\normalsize

\end{itemize}

\subsection{Version SATE-IV.2}

These runs represent using the same SATE2010 training data for \tomcatCase{13} and
\wiresharkCase{0} to test the updated fixed versions (as from SATE2010) to
\tomcatCase{33} and \wiresharkCase{18} using the same settings. At this run,
no new CVEs that may have happened from the previous fixed versions of
\tomcatCase{29} and \wiresharkCase{9} respectively in 2010
were added to the training data for the versions being tested in this experiment
as to see if any old issues reoccur or not. In this short summary, both signal
and NLP testing reveal no same known issues found.

\begin{itemize}
\item
SATE-IV.2-train-test-test-run-quick-tomcat-5-5-33-cve

This is CVE-based classical signal classification.

A typical MARFCAT run. Tomcat 5.5.13 used for training.
For most reports, no warnings were spotted based on what
was learned from 5.5.13, so the reports convey earlier
CVEs were fixed.

Empty reports like:

  \file{report-noprepreprawfftcheb-train-test-test-run-quick-tomcat-5-5-33-cve.xml}

However, the \option{-cos} report is noisy and non-empty:

  \file{report-noprepreprawfftcos-train-test-test-run-quick-tomcat-5-5-33-cve.xml}

Overly detailed log files are also provided.

\item
SATE-IV.2-train-test-test-run-quick-tomcat-5-5-33-cwe

This is classical CWE-based testing.

A typical MARFCAT CWE run. \tomcatCase{13} used for training.

No warnings found based on the CVE data learned.

Most of the reports are empty, e.g.:

  \file{report-nopreprepcharunigramadddelta-train-test-test-run-quick-tomcat-5-5-33-cve-nlp.xml}

The \option{-cos} report is not as noisy as for CVEs, but still contains a couple of false positives.

  \file{report-cweidnoprepreprawfftcos-train-test-test-run-quick-tomcat-5-5-33-cwe.xml}

Overly detailed log files also provided.

Training and testing indexes are provided (\file{*_test.xml} and \file{*_train.xml}).

\item
SATE-IV.2-train-test-test-run-quick-tomcat-5-5-33-cve-nlp

This is CVE-based NLP testing.

A typical MARFCAT NLP run. \tomcatCase{13} used for training.
Usually a slow run, so only one configuration is tried.
No warnings found based on the CVE data learned.

The only empty report is:

  \file{report-nopreprepcharunigramadddelta-train-test-test-run-quick-tomcat-5-5-33-cve-nlp.xml}

However, the \option{-cos} report is noisy and non-empty:

  \file{report-noprepreprawfftcos-train-test-test-run-quick-tomcat-5-5-33-cve.xml}

Overly detailed log files also provided.

Training and testing indexes are provided (\file{*_test.xml} and \file{*_train.xml}).

\item
SATE-IV.2-train-test-test-run-quick-tomcat-5-5-33-cwe-nlp

This is CWE-based NLP testing.

A typical MARFCAT CWE NLP run. \tomcatCase{13} used for training.
Usually a slow run, so only one configuration is tried.
No warnings found based on the CVE data learned.

The only empty report is:

  \file{report-cweidnopreprepcharunigramadddelta-train-test-test-run-quick-tomcat-5-5-33-cwe-nlp.xml}

Overly detailed log files also provided.

\item
SATE-IV.2-train-test-test-run-quick-wireshark-1-2-18-cve

Test \wiresharkCase{18} using the training data from
\wiresharkCase{0} and classical CVE-based processing.

Majority of algorithms returned empty reports.
\option{-cos} was as noisy as usual, but \option{-mink} was non-empty
but quite short (though also presumed with false positives).

Empty reports:

\small
\file{report-noprepreprawfftcheb-train-test-test-run-quick-wireshark-1-2-18-cve.xml}

\file{report-noprepreprawfftdiff-train-test-test-run-quick-wireshark-1-2-18-cve.xml}

\file{report-noprepreprawffteucl-train-test-test-run-quick-wireshark-1-2-18-cve.xml}

\file{report-noprepreprawffthamming-train-test-test-run-quick-wireshark-1-2-18-cve.xml}
\normalsize

Non empty reports: 

\small
\file{report-noprepreprawfftcos-train-test-test-run-quick-wireshark-1-2-18-cve.xml}

\file{report-noprepreprawfftmink-train-test-test-run-quick-wireshark-1-2-18-cve.xml}
\normalsize

\end{itemize}

Verbose log files and input index files are also supplied for the most cases.

{\todo}

\subsection{Version SATE-IV.5}

\subsubsection{Wavelet Experiments}
\label{sect:results-wavelets}

The preliminary experiments using the separating discreet wavelet transform (DWT)
filter are summarized in
\xt{tab:wireshark1-2-0-SATE-IV-5-train-test-run-quick-wireshark-cve-sdwt-spectrogram-graph-flucid}
and \xt{tab:wireshark1-2-0-SATE-IV-5-train-test-run-quick-wireshark-cwe-sdwt-flucid}
for CVEs and CWEs respectively. For comparison, the low-pass FFT filter is used
for the same as shown in
\xt{tab:wireshark1-2-0-SATE-IV-5-train-test-run-quick-wireshark-cve-low-flucid}
and \xt{tab:wireshark1-2-0-SATE-IV-5-train-test-run-quick-wireshark-cwe-low-flucid}
respectively. For the CVE experiments, the wavelet transforms overall produces
better precision across configurations (larger number of configurations produce higher
precision result) than those with the low-pass FFT filter. While the top precision
result remains the same, it is shown than when filtering is wanted, the wavelet
transform is perhaps a better choice for some configurations, e.g., from 4 and below
as well as for the Second Guess statistics.
The very top result for the CWE based processing so far exceeds the overall precision
of separating DWT vs. low-pass FFT, which then drops below for the subsequent configurations.
\option{-cos} was dropped from
\xt{tab:wireshark1-2-0-SATE-IV-5-train-test-run-quick-wireshark-cwe-sdwt-flucid}
for technical reasons. In \xf{fig:spetrogram-wireshark-1-2-0-packet-afs-c-bw-sdwt} is
a spectrogram with the SDWT preprocessing in the pipeline.
More exploration in this area is under way for more advanced wavelet filters
than the simple separating DWT filter as to see whether they would outperform
\option{-raw} or not and at the same time minimizing the run-time performance
decrease with the extra filtering.

\begin{figure}[htpb]
	\begin{center}
	\setlength\fboxsep{0pt}
	\setlength\fboxrule{0.5pt}
	\fbox{\includegraphics[angle=90]{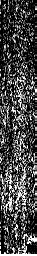}}
	\caption{A spectrogram of \cve{CVE-2009-2562}-vulnerable \file{packet-afs.c} in \wiresharkCase{0}, after SDWT}
	\label{fig:spetrogram-wireshark-1-2-0-packet-afs-c-bw-sdwt}
	\end{center}
\end{figure}

\section{Conclusion}
\label{sect:conclusion}

We review the current results of this experimental work, its current shortcomings,
advantages, and practical implications.

\subsection{Shortcomings}
\label{sect:shortcomings}

Following is a list of the most prominent issues with the presented approach.
Some of them are more ``permanent'', while others are solvable and intended to
be addressed in the future work. Specifically:

\begin{itemize}
\item
Looking at a signal is less intuitive visually for code analysis by humans.
(However, can produce a problematic ``spectrogram'' in some cases).

\item
Line numbers are a problem (easily ``filtered out'' as high-frequency ``noise'', etc.). A whole ``relativistic''
and machine learning methodology developed for the line numbers in \cite{marfcat-arxiv} to compensate for that.
Generally, when CVEs are the primary class, by accurately identifying the CVE number one can get all the
other pertinent details from the CVE database, including patches and line numbers making this a lesser
issue.

\item
Accuracy depends on the quality of the knowledge base (see \xs{sect:kb}) collected.
Some of this collection and annotation is manual to get the indexes right,
and, hence, error prone.
``Garbage in -- garbage out.''

\item
To detect more of the useful CVE or CWE signatures in non-CVE and non-CWE cases requires
large knowledge bases (human-intensive to collect), which can perhaps be shared
by different vendors via a common format, such as SATE, SAFES or {\flucid}.

\item
No path tracing (since no parsing is present); no slicing, semantic annotations,
context, locality of reference, etc. The ``sink'', ``path'', and ``fix'' results
in the reports also have to be machine-learned.

\item
A lot of algorithms and their combinations to try (currently $\approx 1800$ permutations)
to get the best top N. This is, however, also an advantage of the approach as the
underlying framework can quickly allow for such testing.

\item
File-level training vs. fragment-level training -- presently the classes are trained
based on the entire file where weaknesses are found instead of the known file fragments from
CVE-reported patches. The latter would be more fine-grained and precise than whole-file
classification, but slower. However, overall the file-level processing is a man-hour
limitation than a technological one.

\item
Separating wavelet filter performance is rather adversely affects
the precision to low levels.

\item
No nice GUI. Presently the application is script/command-line based.
\end{itemize}

\subsection{Advantages}

There are some key advantages of the approach presented. Some of them
follow:

\begin{itemize}
\item
Relatively fast (e.g., Wireshark's $\approx 2400$ files train and test
in about 3 minutes) on a now-commodity desktop or a laptop.

\item
Language-independent (no parsing) -- given enough examples can apply to any language,
i.e. methodology is the same no matter {\C}, {\cpp}, {\java} or any other source
or binary languages (PHP, C\#, VB, Perl, bytecode, assembly, etc.) are used.

\item
Can automatically learn a large knowledge base to test on known and unknown cases.

\item
Can be used to quickly pre-scan projects for further analysis by humans
or other tools that do in-depth semantic analysis as a means to prioritize.

\item
Can learn from SATE'08, SATE'09, SATE'10, and {\SATE} reports.

\item
Generally, high precision (and recall) in CVE and CWE detection, even at the file level.

\item
A lot of algorithms and their combinations to select the best
for a particular task or class (see \xs{sect:ml-cats}).

\item
Can cope with altered code or code used in other projects
(e.g., a lot of problems in Chrome were found it WebKit, used
by several browsers).
\end{itemize}

\subsection{Practical Implications}

Most practical implications of all static code analyzers are obvious---to
detect source code weaknesses and report them appropriately to
the developers. We outline additional implications this approach brings
to the arsenal below:

\begin{itemize}

\item
The approach can be used on any target language without modifications to the
methodology or knowing the syntax of the language. Thus, it scales to any
popular and new language analysis with a very small amount of effort.

\item
The approach can nearly identically be transposed onto the compiled binaries and bytecode, detecting
vulnerable deployments and installations---akin to virus scanning of
binaries, but instead scanning for infected binaries, one would scan for security-weak binaries
on site deployments to alert system administrators to upgrade their packages.

\item
Can learn from binary signatures from other tools like Snort \cite{snort}.

\item
The approach is easily extendable to the embedded code and mission-critical code
found in aircraft, spacecraft, and various autonomous systems.
\end{itemize}

\subsection{Future Work}

There is a great number of possibilities in the future work. This includes
improvements to the code base of {\marfcat} as well as
resolving unfinished
scenarios and results, addressing shortcomings in \xs{sect:shortcomings},
testing more algorithms and combinations from the related work,
and moving onto other programming languages (e.g., ASP, C\#).
Furthermore, foster collaboration with the academic, industry (such as VeraCode, Coverity) and government vendors
and others who have vast data sets to test the full potential of the approach
with the others and a community as a whole. Then, move on to dynamic code
analysis as well applying similar techniques there.
Other near-future work items include realization of the SVM-based classification,
data export in SAFES and {\flucid} formats, a lot of wavelet filtering improvements,
and distributed {\gipsy} cluster-based evaluation.

To improve detection and classification of the malware in the network traffic or otherwise
we employ machine learning approach to static pcap payload malicious code analysis and fingerprinting
using the open-source {\marf} framework and its {\marfcat} application,
originally designed for the SATE static analysis tool exposition workshop.
We first train on the known malware pcap data and measure the precision and
then test it on the unseen, but known data and select the best available
machine learning combination to do so.
This work elaborates on the details of the methodology
and the corresponding results of application of the machine
learning techniques along with signal processing and NLP alike
to static
network packet
analysis in search for malicious code in the
packet capture (pcap) data.
malicious code analysis
\cite{zeus-botnet-analysis-2010,malware-data-mining,static-anlyzer-save,anti-malware-expert-system,%
pimp-my-pe,%
virus-linguistics,%
novel-adjudification-new-malware,%
auto-classification-inet-malware%
}
We show the system the examples of pcap files with
malware
and {\marfcat} learns them
by computing spectral signatures using signal processing techniques. 
When we test, we compute how similar or distant each file is from the known trained-on malware-laden files.
In part, the methodology can approximately be seen as some signature-based ``antivirus'' or IDS software systems
detect bad signature, except that with a large number of machine learning and signal processing
algorithms, we test to find out which combination gives the highest
precision and best run-time.
At the present, however, we are looking at the whole pcap files.
This aspect lowers the precision, but is fast to scan all the files.
The malware database with known malware, the reports, etc. serves
as a knowledge base to machine-learn from.
Thus, we primarily:
\begin{itemize}
\item Teach the system from the known cases of malware from their pcap data
\item Test on the known cases
\item Test on the unseen cases
\end{itemize}

\subsection{Acknowledgments}

The authors would like to express thanks and gratitude to the following
for their help, resources, advice, and otherwise support and assistance:

\begin{itemize}
	\item NIST SAMATE group
	\item Dr. Brigitte Jaumard
	\item Sleiman Rabah
	\item Open-Source Community
\end{itemize}

\noindent
This work is partially supported by the Faculty of ENCS, Concordia University,
NSERC, and the 2011-2012 CCSEP scholarship.
The wavelet-related work of Yankui Sun is partially supported by the National Natural Science
Foundation of China (No. 60971006).

\appendix

\section{Classification Result Tables}
\label{sect:classification-results}
\label{appdx:classification-results}

What follows are result tables with top classification results
ranked from most precise at the top. This include the configuration
settings for {\marf} by the means of options (the algorithm implementations
are at their defaults \cite{marf-c3s2e08}).

\begin{table}%
\centering
\caption{CVE Stats for \wiresharkCase{0}, Low-Pass FFT Filter Preprocessing}
\label{tab:wireshark1-2-0-SATE-IV-5-train-test-run-quick-wireshark-cve-low-flucid}
\input{results/sate4/wireshark1-2-0-SATE-IV-5-train-test-run-quick-wireshark-cve-low-flucid}
\normalsize
\end{table}

\begin{table}%
\centering
\caption{CVE Stats for \wiresharkCase{0}, Separating DWT Wavelet Filter Preprocessing}
\label{tab:wireshark1-2-0-SATE-IV-5-train-test-run-quick-wireshark-cve-sdwt-spectrogram-graph-flucid}
\input{results/sate4/wireshark1-2-0-SATE-IV-5-train-test-run-quick-wireshark-cve-sdwt-spectrogram-graph-flucid}
\normalsize
\end{table}

\begin{table}%
\centering
\caption{CWE Stats for \wiresharkCase{0}, Separating DWT Wavelet Filter Preprocessing}
\label{tab:wireshark1-2-0-SATE-IV-5-train-test-run-quick-wireshark-cwe-sdwt-flucid}
\input{results/sate4/wireshark1-2-0-SATE-IV-5-train-test-run-quick-wireshark-cwe-sdwt-flucid}
\normalsize
\end{table}

\begin{table}%
\centering
\caption{CWE Stats for \wiresharkCase{0}, Low-Pass FFT Filter Preprocessing}
\label{tab:wireshark1-2-0-SATE-IV-5-train-test-run-quick-wireshark-cwe-low-flucid}
\input{results/sate4/wireshark1-2-0-SATE-IV-5-train-test-run-quick-wireshark-cwe-low-flucid}
\normalsize
\end{table}

\section{{\flucid} Report Example}
\label{sect:flucid-report-example}

An example report encoding the reported data in {\flucid} for
\wiresharkCase{0} after using simple FFT-based feature extraction and
Chebyshev distance as a classifier. The report provides the
same data, compressed, as the SATE XML, but in the {\flucid}
syntax for automated reasoning and event reconstruction during
a digital investigation. The example is a an evidential statement
context encoded for the use in the investigator's knowledge base
of a particular case.

\tiny
\begin{Verbatim}
#FORENSICLUCID
evidential statement report_marfcat_0_0_2_SATE_IV_4
{
	weakness_1 @ [id:1, tool_specific_id:1, cweid:999, cwename:"Insufficient Information (NVD-CWE-noinfo)"]
	where
		dimension id, tool_specific_id, cweid, cwename;
		observation sequence weakness_1 = (locations_wk_1, 1, 0, 1.0);
		locations_wk_1 = locations @ [tool_specific_id:1, cweid:999, cwename:"Insufficient Information (NVD-CWE-noinfo)"];
			observation location_id_340( [line => 828, path => "wireshark-1.2.0/epan/dissectors/packet-afs.c")
			observation location_id_340( [line => 1718, path => "wireshark-1.2.0/epan/dissectors/packet-afs.c")
			observation location_id_340( [line => 1729, path => "wireshark-1.2.0/epan/dissectors/packet-afs.c")
			observation location_id_340( [line => 1740, path => "wireshark-1.2.0/epan/dissectors/packet-afs.c")
			observation location_id_340( [line => 1747, path => "wireshark-1.2.0/epan/dissectors/packet-afs.c")
			textoutput="";
			observation grade = ([ severity => 5, tool_specific_rank => 0.0], 1, 0, 1.0);
	end;
	weakness_2 @ [id:2, tool_specific_id:2, cweid:119, cwename:"Buffer Errors (CWE119)"]
	where
		dimension id, tool_specific_id, cweid, cwename;
		observation sequence weakness_2 = (locations_wk_2, 1, 0, 1.0);
		locations_wk_2 = locations @ [tool_specific_id:2, cweid:119, cwename:"Buffer Errors (CWE119)"];
			observation location_id_411( [line => 830, path => "wireshark-1.2.0/epan/dissectors/packet-ber.c")
			observation location_id_411( [line => 861, path => "wireshark-1.2.0/epan/dissectors/packet-ber.c")
			observation location_id_411( [line => 885, path => "wireshark-1.2.0/epan/dissectors/packet-ber.c")
			textoutput="";
			observation grade = ([ severity => 1, tool_specific_rank => 0.0], 1, 0, 1.0);
	end;
	weakness_3 @ [id:3, tool_specific_id:3, cweid:999, cwename:"Insufficient Information (NVD-CWE-noinfo)"]
	where
		dimension id, tool_specific_id, cweid, cwename;
		observation sequence weakness_3 = (locations_wk_3, 1, 0, 0.004878625561362933);
		locations_wk_3 = locations @ [tool_specific_id:3, cweid:999, cwename:"Insufficient Information (NVD-CWE-noinfo)"];
			observation location_id_433( [line => 669, path => "wireshark-1.2.0/epan/dissectors/packet-btl2cap.c")
			textoutput="";
			observation grade = ([ severity => 5, tool_specific_rank => 204.97576364943077], 1, 0, 0.004878625561362933);
	end;
	weakness_4 @ [id:4, tool_specific_id:4, cweid:999, cwename:"Insufficient Information (NVD-CWE-noinfo)"]
	where
		dimension id, tool_specific_id, cweid, cwename;
		observation sequence weakness_4 = (locations_wk_4, 1, 0, 1.0);
		locations_wk_4 = locations @ [tool_specific_id:4, cweid:999, cwename:"Insufficient Information (NVD-CWE-noinfo)"];
			observation location_id_550( [line => 248, path => "wireshark-1.2.0/epan/dissectors/packet-dcerpc-nt.c")
			observation location_id_550( [line => 252, path => "wireshark-1.2.0/epan/dissectors/packet-dcerpc-nt.c")
			observation location_id_550( [line => 256, path => "wireshark-1.2.0/epan/dissectors/packet-dcerpc-nt.c")
			observation location_id_550( [line => 1138, path => "wireshark-1.2.0/epan/dissectors/packet-dcerpc-nt.c")
			observation location_id_550( [line => 1142, path => "wireshark-1.2.0/epan/dissectors/packet-dcerpc-nt.c")
			observation location_id_550( [line => 1146, path => "wireshark-1.2.0/epan/dissectors/packet-dcerpc-nt.c")
			observation location_id_550( [line => 1201, path => "wireshark-1.2.0/epan/dissectors/packet-dcerpc-nt.c")
			observation location_id_550( [line => 1205, path => "wireshark-1.2.0/epan/dissectors/packet-dcerpc-nt.c")
			observation location_id_550( [line => 1209, path => "wireshark-1.2.0/epan/dissectors/packet-dcerpc-nt.c")
			observation location_id_550( [line => 1314, path => "wireshark-1.2.0/epan/dissectors/packet-dcerpc-nt.c")
			observation location_id_550( [line => 1318, path => "wireshark-1.2.0/epan/dissectors/packet-dcerpc-nt.c")
			observation location_id_550( [line => 1322, path => "wireshark-1.2.0/epan/dissectors/packet-dcerpc-nt.c")
			textoutput="";
			observation grade = ([ severity => 5, tool_specific_rank => 0.0], 1, 0, 1.0);
	end;
	weakness_5 @ [id:5, tool_specific_id:5, cweid:999, cwename:"Insufficient Information (NVD-CWE-noinfo)"]
	where
		dimension id, tool_specific_id, cweid, cwename;
		observation sequence weakness_5 = (locations_wk_5, 1, 0, 0.003778693428627627);
		locations_wk_5 = locations @ [tool_specific_id:5, cweid:999, cwename:"Insufficient Information (NVD-CWE-noinfo)"];
			observation location_id_652( [line => 77, path => "wireshark-1.2.0/epan/dissectors/packet-dtls.c")
			textoutput="";
			observation grade = ([ severity => 5, tool_specific_rank => 264.64173897356557], 1, 0, 0.003778693428627627);
	end;
	weakness_6 @ [id:6, tool_specific_id:6, cweid:999, cwename:"Insufficient Information (NVD-CWE-noinfo)"]
	where
		dimension id, tool_specific_id, cweid, cwename;
		observation sequence weakness_6 = (locations_wk_6, 1, 0, 0.004125022212036806);
		locations_wk_6 = locations @ [tool_specific_id:6, cweid:999, cwename:"Insufficient Information (NVD-CWE-noinfo)"];
			observation location_id_763( [line => 8447, path => "wireshark-1.2.0/epan/dissectors/packet-gsm_a_rr.c")
			textoutput="";
			observation grade = ([ severity => 5, tool_specific_rank => 242.42293704067873], 1, 0, 0.004125022212036806);
	end;
	weakness_7 @ [id:7, tool_specific_id:7, cweid:999, cwename:"Insufficient Information (NVD-CWE-noinfo)"]
	where
		dimension id, tool_specific_id, cweid, cwename;
		observation sequence weakness_7 = (locations_wk_7, 1, 0, 1.0);
		locations_wk_7 = locations @ [tool_specific_id:7, cweid:999, cwename:"Insufficient Information (NVD-CWE-noinfo)"];
			observation location_id_863( [line => 945, path => "wireshark-1.2.0/epan/dissectors/packet-infiniband.c")
			textoutput="";
			observation grade = ([ severity => 5, tool_specific_rank => 0.0], 1, 0, 1.0);
	end;
	weakness_8 @ [id:8, tool_specific_id:8, cweid:119, cwename:"Buffer Errors (CWE119)"]
	where
		dimension id, tool_specific_id, cweid, cwename;
		observation sequence weakness_8 = (locations_wk_8, 1, 0, 1.0);
		locations_wk_8 = locations @ [tool_specific_id:8, cweid:119, cwename:"Buffer Errors (CWE119)"];
			observation location_id_877( [line => 2746, path => "wireshark-1.2.0/epan/dissectors/packet-ipmi-se.c")
			observation location_id_877( [line => 2748, path => "wireshark-1.2.0/epan/dissectors/packet-ipmi-se.c")
			observation location_id_877( [line => 2752, path => "wireshark-1.2.0/epan/dissectors/packet-ipmi-se.c")
			textoutput="";
			observation grade = ([ severity => 1, tool_specific_rank => 0.0], 1, 0, 1.0);
	end;
	weakness_9 @ [id:9, tool_specific_id:9, cweid:998, cwename:"Other (NVD-CWE-Other)"]
	where
		dimension id, tool_specific_id, cweid, cwename;
		observation sequence weakness_9 = (locations_wk_9, 1, 0, 1.0);
		locations_wk_9 = locations @ [tool_specific_id:9, cweid:998, cwename:"Other (NVD-CWE-Other)"];
			observation location_id_882( [line => 792, path => "wireshark-1.2.0/epan/dissectors/packet-ipmi.c")
			textoutput="";
			observation grade = ([ severity => 5, tool_specific_rank => 0.0], 1, 0, 1.0);
	end;
	weakness_10 @ [id:10, tool_specific_id:10, cweid:119, cwename:"Buffer Errors (CWE119)"]
	where
		dimension id, tool_specific_id, cweid, cwename;
		observation sequence weakness_10 = (locations_wk_10, 1, 0, 1.0);
		locations_wk_10 = locations @ [tool_specific_id:10, cweid:119, cwename:"Buffer Errors (CWE119)"];
			observation location_id_969( [line => 523, path => "wireshark-1.2.0/epan/dissectors/packet-lwres.c")
			textoutput="";
			observation grade = ([ severity => 1, tool_specific_rank => 0.0], 1, 0, 1.0);
	end;
	weakness_11 @ [id:11, tool_specific_id:11, cweid:20, cwename:"Input Validation (CWE20)"]
	where
		dimension id, tool_specific_id, cweid, cwename;
		observation sequence weakness_11 = (locations_wk_11, 1, 0, 1.0);
		locations_wk_11 = locations @ [tool_specific_id:11, cweid:20, cwename:"Input Validation (CWE20)"];
			observation location_id_1099( [line => 62, path => "wireshark-1.2.0/epan/dissectors/packet-paltalk.c")
			textoutput="";
			observation grade = ([ severity => 1, tool_specific_rank => 0.0], 1, 0, 1.0);
	end;
	weakness_12 @ [id:12, tool_specific_id:12, cweid:999, cwename:"Insufficient Information (NVD-CWE-noinfo)"]
	where
		dimension id, tool_specific_id, cweid, cwename;
		observation sequence weakness_12 = (locations_wk_12, 1, 0, 0.004878625561362927);
		locations_wk_12 = locations @ [tool_specific_id:12, cweid:999, cwename:"Insufficient Information (NVD-CWE-noinfo)"];
			observation location_id_1174( [line => 897, path => "wireshark-1.2.0/epan/dissectors/packet-radius.c")
			observation location_id_1174( [line => 906, path => "wireshark-1.2.0/epan/dissectors/packet-radius.c")
			observation location_id_1174( [line => 913, path => "wireshark-1.2.0/epan/dissectors/packet-radius.c")
			observation location_id_1174( [line => 1005, path => "wireshark-1.2.0/epan/dissectors/packet-radius.c")
			observation location_id_1174( [line => 1227, path => "wireshark-1.2.0/epan/dissectors/packet-radius.c")
			textoutput="";
			observation grade = ([ severity => 5, tool_specific_rank => 204.975763649431], 1, 0, 0.004878625561362927);
	end;
	weakness_13 @ [id:13, tool_specific_id:13, cweid:999, cwename:"Insufficient Information (NVD-CWE-noinfo)"]
	where
		dimension id, tool_specific_id, cweid, cwename;
		observation sequence weakness_13 = (locations_wk_13, 1, 0, 1.0);
		locations_wk_13 = locations @ [tool_specific_id:13, cweid:999, cwename:"Insufficient Information (NVD-CWE-noinfo)"];
			observation location_id_1282( [line => 1131, path => "wireshark-1.2.0/epan/dissectors/packet-sflow.c")
			textoutput="";
			observation grade = ([ severity => 5, tool_specific_rank => 0.0], 1, 0, 1.0);
	end;
	weakness_14 @ [id:14, tool_specific_id:14, cweid:998, cwename:"Other (NVD-CWE-Other)"]
	where
		dimension id, tool_specific_id, cweid, cwename;
		observation sequence weakness_14 = (locations_wk_14, 1, 0, 1.0);
		locations_wk_14 = locations @ [tool_specific_id:14, cweid:998, cwename:"Other (NVD-CWE-Other)"];
			observation location_id_1303( [line => 2141, path => "wireshark-1.2.0/epan/dissectors/packet-smb-pipe.c")
			textoutput="";
			observation grade = ([ severity => 5, tool_specific_rank => 0.0], 1, 0, 1.0);
	end;
	weakness_15 @ [id:15, tool_specific_id:15, cweid:998, cwename:"Other (NVD-CWE-Other)"]
	where
		dimension id, tool_specific_id, cweid, cwename;
		observation sequence weakness_15 = (locations_wk_15, 1, 0, 1.0);
		locations_wk_15 = locations @ [tool_specific_id:15, cweid:998, cwename:"Other (NVD-CWE-Other)"];
			observation location_id_1307( [line => 8757, path => "wireshark-1.2.0/epan/dissectors/packet-smb.c")
			textoutput="";
			observation grade = ([ severity => 5, tool_specific_rank => 0.0], 1, 0, 1.0);
	end;
	weakness_16 @ [id:16, tool_specific_id:16, cweid:189, cwename:"Numeric Errors (CWE189)"]
	where
		dimension id, tool_specific_id, cweid, cwename;
		observation sequence weakness_16 = (locations_wk_16, 1, 0, 1.0);
		locations_wk_16 = locations @ [tool_specific_id:16, cweid:189, cwename:"Numeric Errors (CWE189)"];
			observation location_id_1307( [line => 2195, path => "wireshark-1.2.0/epan/dissectors/packet-smb.c")
			textoutput="";
			observation grade = ([ severity => 2, tool_specific_rank => 0.0], 1, 0, 1.0);
	end;
	weakness_17 @ [id:17, tool_specific_id:17, cweid:999, cwename:"Insufficient Information (NVD-CWE-noinfo)"]
	where
		dimension id, tool_specific_id, cweid, cwename;
		observation sequence weakness_17 = (locations_wk_17, 1, 0, 0.008328136212759968);
		locations_wk_17 = locations @ [tool_specific_id:17, cweid:999, cwename:"Insufficient Information (NVD-CWE-noinfo)"];
			observation location_id_1307( [line => 8457, path => "wireshark-1.2.0/epan/dissectors/packet-smb.c")
			textoutput="";
			observation grade = ([ severity => 5, tool_specific_rank => 120.07488523877028], 1, 0, 0.008328136212759968);
	end;
	weakness_18 @ [id:18, tool_specific_id:18, cweid:999, cwename:"Insufficient Information (NVD-CWE-noinfo)"]
	where
		dimension id, tool_specific_id, cweid, cwename;
		observation sequence weakness_18 = (locations_wk_18, 1, 0, 0.008328136212759964);
		locations_wk_18 = locations @ [tool_specific_id:18, cweid:999, cwename:"Insufficient Information (NVD-CWE-noinfo)"];
			observation location_id_1309( [line => 955, path => "wireshark-1.2.0/epan/dissectors/packet-smb2.c")
			textoutput="";
			observation grade = ([ severity => 5, tool_specific_rank => 120.07488523877032], 1, 0, 0.008328136212759964);
	end;
	weakness_19 @ [id:19, tool_specific_id:19, cweid:999, cwename:"Insufficient Information (NVD-CWE-noinfo)"]
	where
		dimension id, tool_specific_id, cweid, cwename;
		observation sequence weakness_19 = (locations_wk_19, 1, 0, 0.004321352067642762);
		locations_wk_19 = locations @ [tool_specific_id:19, cweid:999, cwename:"Insufficient Information (NVD-CWE-noinfo)"];
			observation location_id_1333( [line => 813, path => "wireshark-1.2.0/epan/dissectors/packet-ssl-utils.c")
			observation location_id_1333( [line => 843, path => "wireshark-1.2.0/epan/dissectors/packet-ssl-utils.c")
			textoutput="";
			observation grade = ([ severity => 5, tool_specific_rank => 231.40905539443497], 1, 0, 0.004321352067642762);
	end;
	weakness_20 @ [id:20, tool_specific_id:20, cweid:999, cwename:"Insufficient Information (NVD-CWE-noinfo)"]
	where
		dimension id, tool_specific_id, cweid, cwename;
		observation sequence weakness_20 = (locations_wk_20, 1, 0, 0.0021114804997331383);
		locations_wk_20 = locations @ [tool_specific_id:20, cweid:999, cwename:"Insufficient Information (NVD-CWE-noinfo)"];
			observation location_id_1334( [line => 153, path => "wireshark-1.2.0/epan/dissectors/packet-ssl-utils.h")
			textoutput="";
			observation grade = ([ severity => 5, tool_specific_rank => 473.60134281438354], 1, 0, 0.0021114804997331383);
	end;
	weakness_21 @ [id:21, tool_specific_id:21, cweid:999, cwename:"Insufficient Information (NVD-CWE-noinfo)"]
	where
		dimension id, tool_specific_id, cweid, cwename;
		observation sequence weakness_21 = (locations_wk_21, 1, 0, 0.003463630817441021);
		locations_wk_21 = locations @ [tool_specific_id:21, cweid:999, cwename:"Insufficient Information (NVD-CWE-noinfo)"];
			observation location_id_1335( [line => 275, path => "wireshark-1.2.0/epan/dissectors/packet-ssl.c")
			textoutput="";
			observation grade = ([ severity => 5, tool_specific_rank => 288.71437306901373], 1, 0, 0.003463630817441021);
	end;
	weakness_22 @ [id:22, tool_specific_id:22, cweid:999, cwename:"Insufficient Information (NVD-CWE-noinfo)"]
	where
		dimension id, tool_specific_id, cweid, cwename;
		observation sequence weakness_22 = (locations_wk_22, 1, 0, 0.004125022212036806);
		locations_wk_22 = locations @ [tool_specific_id:22, cweid:999, cwename:"Insufficient Information (NVD-CWE-noinfo)"];
			observation location_id_1583( [line => 1799, path => "wireshark-1.2.0/epan/packet.c")
			textoutput="";
			observation grade = ([ severity => 5, tool_specific_rank => 242.42293704067873], 1, 0, 0.004125022212036806);
	end;
	weakness_23 @ [id:23, tool_specific_id:23, cweid:399, cwename:"Resource Management Errors (CWE399)"]
	where
		dimension id, tool_specific_id, cweid, cwename;
		observation sequence weakness_23 = (locations_wk_23, 1, 0, 1.0);
		locations_wk_23 = locations @ [tool_specific_id:23, cweid:399, cwename:"Resource Management Errors (CWE399)"];
			observation location_id_1611( [line => 345, path => "wireshark-1.2.0/epan/sigcomp-udvm.c")
			textoutput="";
			observation grade = ([ severity => 3, tool_specific_rank => 0.0], 1, 0, 1.0);
	end;
	weakness_24 @ [id:24, tool_specific_id:24, cweid:119, cwename:"Buffer Errors (CWE119)"]
	where
		dimension id, tool_specific_id, cweid, cwename;
		observation sequence weakness_24 = (locations_wk_24, 1, 0, 1.0);
		locations_wk_24 = locations @ [tool_specific_id:24, cweid:119, cwename:"Buffer Errors (CWE119)"];
			observation location_id_1611( [line => 321, path => "wireshark-1.2.0/epan/sigcomp-udvm.c")
			textoutput="";
			observation grade = ([ severity => 1, tool_specific_rank => 0.0], 1, 0, 1.0);
	end;
	weakness_25 @ [id:25, tool_specific_id:25, cweid:20, cwename:"Input Validation (CWE20)"]
	where
		dimension id, tool_specific_id, cweid, cwename;
		observation sequence weakness_25 = (locations_wk_25, 1, 0, 0.001495003320843141);
		locations_wk_25 = locations @ [tool_specific_id:25, cweid:20, cwename:"Input Validation (CWE20)"];
			observation location_id_2012( [line => 89, path => "wireshark-1.2.0/plugins/docsis/packet-bpkmreq.c")
			textoutput="";
			observation grade = ([ severity => 1, tool_specific_rank => 668.8948352542972], 1, 0, 0.001495003320843141);
	end;
	weakness_26 @ [id:26, tool_specific_id:26, cweid:20, cwename:"Input Validation (CWE20)"]
	where
		dimension id, tool_specific_id, cweid, cwename;
		observation sequence weakness_26 = (locations_wk_26, 1, 0, 0.0014959114047375394);
		locations_wk_26 = locations @ [tool_specific_id:26, cweid:20, cwename:"Input Validation (CWE20)"];
			observation location_id_2013( [line => 90, path => "wireshark-1.2.0/plugins/docsis/packet-bpkmrsp.c")
			textoutput="";
			observation grade = ([ severity => 1, tool_specific_rank => 668.4887867242726], 1, 0, 0.0014959114047375394);
	end;
	weakness_27 @ [id:27, tool_specific_id:27, cweid:20, cwename:"Input Validation (CWE20)"]
	where
		dimension id, tool_specific_id, cweid, cwename;
		observation sequence weakness_27 = (locations_wk_27, 1, 0, 0.002153585613826869);
		locations_wk_27 = locations @ [tool_specific_id:27, cweid:20, cwename:"Input Validation (CWE20)"];
			observation location_id_2020( [line => 72, path => "wireshark-1.2.0/plugins/docsis/packet-dsaack.c")
			textoutput="";
			observation grade = ([ severity => 1, tool_specific_rank => 464.341883405798], 1, 0, 0.002153585613826869);
	end;
	weakness_28 @ [id:28, tool_specific_id:28, cweid:20, cwename:"Input Validation (CWE20)"]
	where
		dimension id, tool_specific_id, cweid, cwename;
		observation sequence weakness_28 = (locations_wk_28, 1, 0, 0.00229165238295895);
		locations_wk_28 = locations @ [tool_specific_id:28, cweid:20, cwename:"Input Validation (CWE20)"];
			observation location_id_2022( [line => 72, path => "wireshark-1.2.0/plugins/docsis/packet-dsarsp.c")
			textoutput="";
			observation grade = ([ severity => 1, tool_specific_rank => 436.36635618741343], 1, 0, 0.00229165238295895);
	end;
	weakness_29 @ [id:29, tool_specific_id:29, cweid:20, cwename:"Input Validation (CWE20)"]
	where
		dimension id, tool_specific_id, cweid, cwename;
		observation sequence weakness_29 = (locations_wk_29, 1, 0, 0.002184230355798278);
		locations_wk_29 = locations @ [tool_specific_id:29, cweid:20, cwename:"Input Validation (CWE20)"];
			observation location_id_2023( [line => 72, path => "wireshark-1.2.0/plugins/docsis/packet-dscack.c")
			textoutput="";
			observation grade = ([ severity => 1, tool_specific_rank => 457.82716889058463], 1, 0, 0.002184230355798278);
	end;
	weakness_30 @ [id:30, tool_specific_id:30, cweid:20, cwename:"Input Validation (CWE20)"]
	where
		dimension id, tool_specific_id, cweid, cwename;
		observation sequence weakness_30 = (locations_wk_30, 1, 0, 0.0023006295251237975);
		locations_wk_30 = locations @ [tool_specific_id:30, cweid:20, cwename:"Input Validation (CWE20)"];
			observation location_id_2025( [line => 73, path => "wireshark-1.2.0/plugins/docsis/packet-dscrsp.c")
			textoutput="";
			observation grade = ([ severity => 1, tool_specific_rank => 434.6636383996635], 1, 0, 0.0023006295251237975);
	end;
	weakness_31 @ [id:31, tool_specific_id:31, cweid:20, cwename:"Input Validation (CWE20)"]
	where
		dimension id, tool_specific_id, cweid, cwename;
		observation sequence weakness_31 = (locations_wk_31, 1, 0, 0.001897888480826156);
		locations_wk_31 = locations @ [tool_specific_id:31, cweid:20, cwename:"Input Validation (CWE20)"];
			observation location_id_2032( [line => 72, path => "wireshark-1.2.0/plugins/docsis/packet-regack.c")
			textoutput="";
			observation grade = ([ severity => 1, tool_specific_rank => 526.9013485790784], 1, 0, 0.001897888480826156);
	end;
	weakness_32 @ [id:32, tool_specific_id:32, cweid:20, cwename:"Input Validation (CWE20)"]
	where
		dimension id, tool_specific_id, cweid, cwename;
		observation sequence weakness_32 = (locations_wk_32, 1, 0, 0.002216818195096963);
		locations_wk_32 = locations @ [tool_specific_id:32, cweid:20, cwename:"Input Validation (CWE20)"];
			observation location_id_2035( [line => 73, path => "wireshark-1.2.0/plugins/docsis/packet-regrsp.c")
			textoutput="";
			observation grade = ([ severity => 1, tool_specific_rank => 451.096983149879], 1, 0, 0.002216818195096963);
	end;
	weakness_33 @ [id:33, tool_specific_id:33, cweid:999, cwename:"Insufficient Information (NVD-CWE-noinfo)"]
	where
		dimension id, tool_specific_id, cweid, cwename;
		observation sequence weakness_33 = (locations_wk_33, 1, 0, 0.0028814675905206645);
		locations_wk_33 = locations @ [tool_specific_id:33, cweid:999, cwename:"Insufficient Information (NVD-CWE-noinfo)"];
			observation location_id_2097( [line => 433, path => "wireshark-1.2.0/plugins/opcua/opcua_complextypeparser.c")
			textoutput="";
			observation grade = ([ severity => 5, tool_specific_rank => 347.04537482557834], 1, 0, 0.0028814675905206645);
	end;
	weakness_34 @ [id:34, tool_specific_id:34, cweid:999, cwename:"Insufficient Information (NVD-CWE-noinfo)"]
	where
		dimension id, tool_specific_id, cweid, cwename;
		observation sequence weakness_34 = (locations_wk_34, 1, 0, 0.0028288900371324934);
		locations_wk_34 = locations @ [tool_specific_id:34, cweid:999, cwename:"Insufficient Information (NVD-CWE-noinfo)"];
			observation location_id_2107( [line => 616, path => "wireshark-1.2.0/plugins/opcua/opcua_serviceparser.c")
			textoutput="";
			observation grade = ([ severity => 5, tool_specific_rank => 353.49553601371184], 1, 0, 0.0028288900371324934);
	end;
	weakness_35 @ [id:35, tool_specific_id:35, cweid:999, cwename:"Insufficient Information (NVD-CWE-noinfo)"]
	where
		dimension id, tool_specific_id, cweid, cwename;
		observation sequence weakness_35 = (locations_wk_35, 1, 0, 0.003058220966230374);
		locations_wk_35 = locations @ [tool_specific_id:35, cweid:999, cwename:"Insufficient Information (NVD-CWE-noinfo)"];
			observation location_id_2110( [line => 340, path => "wireshark-1.2.0/plugins/opcua/opcua_simpletypes.c")
			textoutput="";
			observation grade = ([ severity => 5, tool_specific_rank => 326.9874907805045], 1, 0, 0.003058220966230374);
	end;
	weakness_36 @ [id:36, tool_specific_id:36, cweid:999, cwename:"Insufficient Information (NVD-CWE-noinfo)"]
	where
		dimension id, tool_specific_id, cweid, cwename;
		observation sequence weakness_36 = (locations_wk_36, 1, 0, 0.0018096494904338023);
		locations_wk_36 = locations @ [tool_specific_id:36, cweid:999, cwename:"Insufficient Information (NVD-CWE-noinfo)"];
			observation location_id_2112( [line => 132, path => "wireshark-1.2.0/plugins/opcua/opcua_transport_layer.c")
			observation location_id_2112( [line => 169, path => "wireshark-1.2.0/plugins/opcua/opcua_transport_layer.c")
			observation location_id_2112( [line => 181, path => "wireshark-1.2.0/plugins/opcua/opcua_transport_layer.c")
			observation location_id_2112( [line => 195, path => "wireshark-1.2.0/plugins/opcua/opcua_transport_layer.c")
			observation location_id_2112( [line => 226, path => "wireshark-1.2.0/plugins/opcua/opcua_transport_layer.c")
			observation location_id_2112( [line => 250, path => "wireshark-1.2.0/plugins/opcua/opcua_transport_layer.c")
			textoutput="";
			observation grade = ([ severity => 5, tool_specific_rank => 552.593198454295], 1, 0, 0.0018096494904338023);
	end;
	weakness_37 @ [id:37, tool_specific_id:37, cweid:119, cwename:"Buffer Errors (CWE119)"]
	where
		dimension id, tool_specific_id, cweid, cwename;
		observation sequence weakness_37 = (locations_wk_37, 1, 0, 1.0);
		locations_wk_37 = locations @ [tool_specific_id:37, cweid:119, cwename:"Buffer Errors (CWE119)"];
			observation location_id_2321( [line => 149, path => "wireshark-1.2.0/wiretap/daintree-sna.c")
			observation location_id_2321( [line => 205, path => "wireshark-1.2.0/wiretap/daintree-sna.c")
			textoutput="";
			observation grade = ([ severity => 1, tool_specific_rank => 0.0], 1, 0, 1.0);
	end;
	weakness_38 @ [id:38, tool_specific_id:38, cweid:189, cwename:"Numeric Errors (CWE189)"]
	where
		dimension id, tool_specific_id, cweid, cwename;
		observation sequence weakness_38 = (locations_wk_38, 1, 0, 1.0);
		locations_wk_38 = locations @ [tool_specific_id:38, cweid:189, cwename:"Numeric Errors (CWE189)"];
			observation location_id_2327( [line => 228, path => "wireshark-1.2.0/wiretap/erf.c")
			textoutput="";
			observation grade = ([ severity => 2, tool_specific_rank => 0.0], 1, 0, 1.0);
	end;
}
\end{Verbatim}
\normalsize

\bibliographystyle{alpha}
\NIST{}
\bibliography{marf-sate}

\clearpage
\let\oldtwocolumn\twocolumn
\renewcommand{\twocolumn}[1][]{#1}
\printindex
\renewcommand{\twocolumn}[1][]{\oldtwocolumn}

\end{document}

%% file: commands.tex
\newcommand{\xf}[1]{Figure~\ref{#1}}

\newcommand{\xs}[1]{Section~\ref{#1}}

\newcommand{\xt}[1]{Table~\ref{#1}}

\newcommand{\jini}{{Jini\index{Jini}}}
\newcommand{\jms}{{JMS\index{JMS}}}

\newcommand{\gipsy}{{GIPSY\index{GIPSY}}}

\newcommand{\flucid}{{Forensic Lucid\index{Forensic Lucid}}}

\newcommand{\C}{{C\index{C}}}
\newcommand{\cpp}{{C++\index{C++}}}
\newcommand{\perl}{{Perl\index{Perl}}}
\newcommand{\java}{{Java\index{Java}}}

\newcommand{\php}{{PHP\index{PHP}}}

\newcommand{\todo}[0]
{
	\begin{center}{\Large [TODO]}\index{TODO}\end{center}
}

\newcommand{\file}[1]{\url{#1}\index{Files!\url{#1}}}
\newcommand{\tool}[1]{\texttt{#1}\index{Tools!#1}}
\newcommand{\option}[1]{\texttt{#1}\index{Options!#1}}
\newcommand{\api}[1]{\texttt{#1}\index{API!#1}}

\newcommand{\marf}[0]{MARF\index{MARF}\index{Frameworks!MARF}\index{Libraries!MARF}}

\newcommand{\lucidL}[1]{{$\mathit{Lucid}$}($L$) }

		{}

\def\myvert{\raise 2.27pt \hbox{\vrule depth 0pt height 8pt width 0.2mm}}
\def\myarrow{\hspace*{0.43mm}%
             \raise 2.29pt\hbox{\vrule depth 0pt height 8pt width 0.16mm}%
             \hspace*{-0.32mm}%
             $\longrightarrow$
             \ %
             }


%% file: results/sate4/wireshark120-cwe-half-training-full-testing.tex
\begin{tabular}{|c|r|l|c|c|r|}\hline
guess & run & algorithms & good & bad &  \% \\\hline
1st & 1 & \option{-cweid} \option{-nopreprep} \option{-raw} \option{-fft} \option{-cheb}  & 26 & 10 & 72.22\\\hline
1st & 2 & \option{-cweid} \option{-nopreprep} \option{-raw} \option{-fft} \option{-diff}  & 26 & 10 & 72.22\\\hline
1st & 3 & \option{-cweid} \option{-nopreprep} \option{-raw} \option{-fft} \option{-eucl}  & 22 & 14 & 61.11\\\hline
1st & 4 & \option{-cweid} \option{-nopreprep} \option{-raw} \option{-fft} \option{-cos}  & 25 & 23 & 52.08\\\hline
1st & 5 & \option{-cweid} \option{-nopreprep} \option{-raw} \option{-fft} \option{-mink}  & 17 & 19 & 47.22\\\hline
1st & 6 & \option{-cweid} \option{-nopreprep} \option{-raw} \option{-fft} \option{-hamming}  & 17 & 19 & 47.22\\\hline
2nd & 1 & \option{-cweid} \option{-nopreprep} \option{-raw} \option{-fft} \option{-cheb}  & 30 & 6 & 83.33\\\hline
2nd & 2 & \option{-cweid} \option{-nopreprep} \option{-raw} \option{-fft} \option{-diff}  & 30 & 6 & 83.33\\\hline
2nd & 3 & \option{-cweid} \option{-nopreprep} \option{-raw} \option{-fft} \option{-eucl}  & 24 & 12 & 66.67\\\hline
2nd & 4 & \option{-cweid} \option{-nopreprep} \option{-raw} \option{-fft} \option{-cos}  & 32 & 16 & 66.67\\\hline
2nd & 5 & \option{-cweid} \option{-nopreprep} \option{-raw} \option{-fft} \option{-mink}  & 23 & 13 & 63.89\\\hline
2nd & 6 & \option{-cweid} \option{-nopreprep} \option{-raw} \option{-fft} \option{-hamming}  & 24 & 12 & 66.67\\\hline
guess & run & class & good & bad &  \% \\\hline
1st & 1 & \cwe{NVD-CWE-noinfo} & 68 & 39 & 63.55\\\hline
1st & 2 & \cwe{CWE-20} & 38 & 22 & 63.33\\\hline
1st & 3 & \cwe{CWE-119} & 18 & 14 & 56.25\\\hline
1st & 4 & \cwe{NVD-CWE-Other} & 9 & 8 & 52.94\\\hline
1st & 5 & \cwe{CWE-189} & 0 & 12 & 0.00\\\hline
2nd & 1 & \cwe{NVD-CWE-noinfo} & 84 & 23 & 78.50\\\hline
2nd & 2 & \cwe{CWE-20} & 39 & 21 & 65.00\\\hline
2nd & 3 & \cwe{CWE-119} & 29 & 3 & 90.62\\\hline
2nd & 4 & \cwe{NVD-CWE-Other} & 11 & 6 & 64.71\\\hline
2nd & 5 & \cwe{CWE-189} & 0 & 12 & 0.00\\\hline
\end{tabular}

%% file: results/sate4/tomcat13-cwe-half-train-full-test.tex
\begin{tabular}{|c|r|l|c|c|r|}\hline
guess & run & algorithms & good & bad &  \% \\\hline
1st & 1 & \option{-cweid} \option{-nopreprep} \option{-raw} \option{-fft} \option{-diff}  & 6 & 2 & 75.00\\\hline
1st & 2 & \option{-cweid} \option{-nopreprep} \option{-raw} \option{-fft} \option{-hamming}  & 5 & 9 & 35.71\\\hline
2nd & 1 & \option{-cweid} \option{-nopreprep} \option{-raw} \option{-fft} \option{-diff}  & 6 & 2 & 75.00\\\hline
2nd & 2 & \option{-cweid} \option{-nopreprep} \option{-raw} \option{-fft} \option{-hamming}  & 8 & 6 & 57.14\\\hline
guess & run & class & good & bad &  \% \\\hline
1st & 1 & \cwe{CWE-264} & 1 & 0 & 100.00\\\hline
1st & 2 & \cwe{CWE-255} & 2 & 0 & 100.00\\\hline
1st & 3 & \cwe{CWE-200} & 1 & 0 & 100.00\\\hline
1st & 4 & \cwe{CWE-22} & 6 & 3 & 66.67\\\hline
1st & 5 & \cwe{CWE-79} & 1 & 4 & 20.00\\\hline
1st & 6 & \cwe{CWE-119} & 0 & 2 & 0.00\\\hline
1st & 7 & \cwe{CWE-20} & 0 & 2 & 0.00\\\hline
2nd & 1 & \cwe{CWE-264} & 1 & 0 & 100.00\\\hline
2nd & 2 & \cwe{CWE-255} & 2 & 0 & 100.00\\\hline
2nd & 3 & \cwe{CWE-200} & 1 & 0 & 100.00\\\hline
2nd & 4 & \cwe{CWE-22} & 7 & 2 & 77.78\\\hline
2nd & 5 & \cwe{CWE-79} & 3 & 2 & 60.00\\\hline
2nd & 6 & \cwe{CWE-119} & 0 & 2 & 0.00\\\hline
2nd & 7 & \cwe{CWE-20} & 0 & 2 & 0.00\\\hline
\end{tabular}

%% file: results/sate4/chrome54-cwe-half-train-full-test.tex
\begin{tabular}{|c|r|l|c|c|r|}\hline
guess & run & algorithms & good & bad &  \% \\\hline
1st & 1 & \option{-cweid} \option{-nopreprep} \option{-raw} \option{-fft} \option{-cos}  & 2 & 0 & 100.00\\\hline
1st & 2 & \option{-cweid} \option{-nopreprep} \option{-raw} \option{-fft} \option{-eucl}  & 4 & 5 & 44.44\\\hline
1st & 3 & \option{-cweid} \option{-nopreprep} \option{-raw} \option{-fft} \option{-cheb}  & 3 & 6 & 33.33\\\hline
1st & 4 & \option{-cweid} \option{-nopreprep} \option{-raw} \option{-fft} \option{-hamming}  & 3 & 6 & 33.33\\\hline
1st & 5 & \option{-cweid} \option{-nopreprep} \option{-raw} \option{-fft} \option{-mink}  & 2 & 7 & 22.22\\\hline
2nd & 1 & \option{-cweid} \option{-nopreprep} \option{-raw} \option{-fft} \option{-cos}  & 2 & 0 & 100.00\\\hline
2nd & 2 & \option{-cweid} \option{-nopreprep} \option{-raw} \option{-fft} \option{-eucl}  & 4 & 5 & 44.44\\\hline
2nd & 3 & \option{-cweid} \option{-nopreprep} \option{-raw} \option{-fft} \option{-cheb}  & 4 & 5 & 44.44\\\hline
2nd & 4 & \option{-cweid} \option{-nopreprep} \option{-raw} \option{-fft} \option{-hamming}  & 4 & 5 & 44.44\\\hline
2nd & 5 & \option{-cweid} \option{-nopreprep} \option{-raw} \option{-fft} \option{-mink}  & 3 & 6 & 33.33\\\hline
guess & run & class & good & bad &  \% \\\hline
1st & 1 & \cwe{CWE-94} & 6 & 3 & 66.67\\\hline
1st & 2 & \cwe{CWE-20} & 3 & 2 & 60.00\\\hline
1st & 3 & \cwe{CWE-79} & 2 & 2 & 50.00\\\hline
1st & 4 & \cwe{NVD-CWE-noinfo} & 2 & 2 & 50.00\\\hline
1st & 5 & \cwe{NVD-CWE-Other} & 1 & 7 & 12.50\\\hline
1st & 6 & \cwe{CWE-399} & 0 & 4 & 0.00\\\hline
1st & 7 & \cwe{CWE-119} & 0 & 4 & 0.00\\\hline
2nd & 1 & \cwe{CWE-94} & 6 & 3 & 66.67\\\hline
2nd & 2 & \cwe{CWE-20} & 3 & 2 & 60.00\\\hline
2nd & 3 & \cwe{CWE-79} & 3 & 1 & 75.00\\\hline
2nd & 4 & \cwe{NVD-CWE-noinfo} & 3 & 1 & 75.00\\\hline
2nd & 5 & \cwe{NVD-CWE-Other} & 2 & 6 & 25.00\\\hline
2nd & 6 & \cwe{CWE-399} & 0 & 4 & 0.00\\\hline
2nd & 7 & \cwe{CWE-119} & 0 & 4 & 0.00\\\hline
\end{tabular}

%% file: results/sate4/wireshark120-cwe-half-training-full-testing-nlp.tex
\begin{tabular}{|c|r|l|c|c|r|}\hline
guess & run & algorithms & good & bad &  \% \\\hline
1st & 1 & \option{-cweid} \option{-nopreprep} \option{-char} \option{-unigram} -add-delta  & 15 & 21 & 41.67\\\hline
2nd & 1 & \option{-cweid} \option{-nopreprep} \option{-char} \option{-unigram} -add-delta  & 23 & 13 & 63.89\\\hline
guess & run & class & good & bad &  \% \\\hline
1st & 1 & \cwe{NVD-CWE-noinfo} & 11 & 7 & 61.11\\\hline
1st & 2 & \cwe{NVD-CWE-Other} & 1 & 1 & 50.00\\\hline
1st & 3 & \cwe{CWE-119} & 2 & 3 & 40.00\\\hline
1st & 4 & \cwe{CWE-20} & 1 & 9 & 10.00\\\hline
1st & 5 & \cwe{CWE-189} & 0 & 1 & 0.00\\\hline
2nd & 1 & \cwe{NVD-CWE-noinfo} & 17 & 1 & 94.44\\\hline
2nd & 2 & \cwe{NVD-CWE-Other} & 1 & 1 & 50.00\\\hline
2nd & 3 & \cwe{CWE-119} & 4 & 1 & 80.00\\\hline
2nd & 4 & \cwe{CWE-20} & 1 & 9 & 10.00\\\hline
2nd & 5 & \cwe{CWE-189} & 0 & 1 & 0.00\\\hline
\end{tabular}

%% file: results/sate4/tomcat13-cwe-half-train-full-test-nlp.tex
\begin{tabular}{|c|r|l|c|c|r|}\hline
guess & run & algorithms & good & bad &  \% \\\hline
1st & 1 & \option{-cweid} \option{-nopreprep} \option{-char} \option{-unigram} -add-delta  & 14 & 19 & 42.42\\\hline
2nd & 1 & \option{-cweid} \option{-nopreprep} \option{-char} \option{-unigram} -add-delta  & 18 & 15 & 54.55\\\hline
guess & run & class & good & bad &  \% \\\hline
1st & 1 & \cwe{CWE-255} & 1 & 0 & 100.00\\\hline
1st & 2 & \cwe{CWE-264} & 2 & 0 & 100.00\\\hline
1st & 3 & \cwe{CWE-119} & 1 & 0 & 100.00\\\hline
1st & 4 & \cwe{CWE-20} & 1 & 0 & 100.00\\\hline
1st & 5 & \cwe{CWE-22} & 7 & 9 & 43.75\\\hline
1st & 6 & \cwe{CWE-200} & 1 & 3 & 25.00\\\hline
1st & 7 & \cwe{CWE-79} & 1 & 6 & 14.29\\\hline
1st & 8 & \cwe{CWE-16} & 0 & 1 & 0.00\\\hline
2nd & 1 & \cwe{CWE-255} & 1 & 0 & 100.00\\\hline
2nd & 2 & \cwe{CWE-264} & 2 & 0 & 100.00\\\hline
2nd & 3 & \cwe{CWE-119} & 1 & 0 & 100.00\\\hline
2nd & 4 & \cwe{CWE-20} & 1 & 0 & 100.00\\\hline
2nd & 5 & \cwe{CWE-22} & 11 & 5 & 68.75\\\hline
2nd & 6 & \cwe{CWE-200} & 1 & 3 & 25.00\\\hline
2nd & 7 & \cwe{CWE-79} & 1 & 6 & 14.29\\\hline
2nd & 8 & \cwe{CWE-16} & 0 & 1 & 0.00\\\hline
\end{tabular}

%% file: results/sate4/chrome54-cwe-half-train-full-test-nlp.tex
\begin{tabular}{|c|r|l|c|c|r|}\hline
guess & run & algorithms & good & bad &  \% \\\hline
1st & 1 & \option{-cweid} \option{-nopreprep} \option{-char} \option{-unigram} -add-delta  & 4 & 5 & 44.44\\\hline
2nd & 1 & \option{-cweid} \option{-nopreprep} \option{-char} \option{-unigram} -add-delta  & 5 & 4 & 55.56\\\hline
guess & run & class & good & bad &  \% \\\hline
1st & 1 & \cwe{NVD-CWE-noinfo} & 1 & 0 & 100.00\\\hline
1st & 2 & \cwe{CWE-79} & 1 & 0 & 100.00\\\hline
1st & 3 & \cwe{CWE-20} & 1 & 0 & 100.00\\\hline
1st & 4 & \cwe{CWE-94} & 1 & 1 & 50.00\\\hline
1st & 5 & \cwe{CWE-399} & 0 & 1 & 0.00\\\hline
1st & 6 & \cwe{NVD-CWE-Other} & 0 & 2 & 0.00\\\hline
1st & 7 & \cwe{CWE-119} & 0 & 1 & 0.00\\\hline
2nd & 1 & \cwe{NVD-CWE-noinfo} & 1 & 0 & 100.00\\\hline
2nd & 2 & \cwe{CWE-79} & 1 & 0 & 100.00\\\hline
2nd & 3 & \cwe{CWE-20} & 1 & 0 & 100.00\\\hline
2nd & 4 & \cwe{CWE-94} & 1 & 1 & 50.00\\\hline
2nd & 5 & \cwe{CWE-399} & 0 & 1 & 0.00\\\hline
2nd & 6 & \cwe{NVD-CWE-Other} & 0 & 2 & 0.00\\\hline
2nd & 7 & \cwe{CWE-119} & 1 & 0 & 100.00\\\hline
\end{tabular}

%% file: results/sate4/wireshark1-2-0-SATE-IV-5-train-test-run-quick-wireshark-cve-low-flucid.tex
\begin{tabular}{|c|r|l|c|c|r|}\hline
guess & run & algorithms & good & bad &  \% \\\hline
1st & 1 & \option{-nopreprep} \option{-low} \option{-fft} \option{-cheb} \option{-flucid}  & 37 & 4 & 90.24\\\hline
1st & 2 & \option{-nopreprep} \option{-low} \option{-fft} \option{-diff} \option{-flucid}  & 37 & 4 & 90.24\\\hline
1st & 3 & \option{-nopreprep} \option{-low} \option{-fft} \option{-eucl} \option{-flucid}  & 27 & 14 & 65.85\\\hline
1st & 4 & \option{-nopreprep} \option{-low} \option{-fft} \option{-hamming} \option{-flucid}  & 23 & 18 & 56.10\\\hline
1st & 5 & \option{-nopreprep} \option{-low} \option{-fft} \option{-mink} \option{-flucid}  & 22 & 19 & 53.66\\\hline
1st & 6 & \option{-nopreprep} \option{-low} \option{-fft} \option{-cos} \option{-flucid}  & 36 & 114 & 24.00\\\hline
2nd & 1 & \option{-nopreprep} \option{-low} \option{-fft} \option{-cheb} \option{-flucid}  & 38 & 3 & 92.68\\\hline
2nd & 2 & \option{-nopreprep} \option{-low} \option{-fft} \option{-diff} \option{-flucid}  & 38 & 3 & 92.68\\\hline
2nd & 3 & \option{-nopreprep} \option{-low} \option{-fft} \option{-eucl} \option{-flucid}  & 34 & 7 & 82.93\\\hline
2nd & 4 & \option{-nopreprep} \option{-low} \option{-fft} \option{-hamming} \option{-flucid}  & 26 & 15 & 63.41\\\hline
2nd & 5 & \option{-nopreprep} \option{-low} \option{-fft} \option{-mink} \option{-flucid}  & 31 & 10 & 75.61\\\hline
2nd & 6 & \option{-nopreprep} \option{-low} \option{-fft} \option{-cos} \option{-flucid}  & 39 & 111 & 26.00\\\hline
guess & run & class & good & bad &  \% \\\hline
1st & 1 & \cve{CVE-2009-3829} & 6 & 0 & 100.00\\\hline
1st & 2 & \cve{CVE-2009-4376} & 6 & 0 & 100.00\\\hline
1st & 3 & \cve{CVE-2010-0304} & 6 & 0 & 100.00\\\hline
1st & 4 & \cve{CVE-2010-2286} & 6 & 0 & 100.00\\\hline
1st & 5 & \cve{CVE-2010-2283} & 6 & 0 & 100.00\\\hline
1st & 6 & \cve{CVE-2009-3551} & 6 & 0 & 100.00\\\hline
1st & 7 & \cve{CVE-2009-3549} & 6 & 0 & 100.00\\\hline
1st & 8 & \cve{CVE-2009-3241} & 15 & 9 & 62.50\\\hline
1st & 9 & \cve{CVE-2009-2560} & 9 & 6 & 60.00\\\hline
1st & 10 & \cve{CVE-2010-1455} & 30 & 24 & 55.56\\\hline
1st & 11 & \cve{CVE-2009-2563} & 6 & 5 & 54.55\\\hline
1st & 12 & \cve{CVE-2009-2562} & 6 & 5 & 54.55\\\hline
1st & 13 & \cve{CVE-2009-2561} & 6 & 7 & 46.15\\\hline
1st & 14 & \cve{CVE-2009-4378} & 6 & 7 & 46.15\\\hline
1st & 15 & \cve{CVE-2010-2287} & 6 & 7 & 46.15\\\hline
1st & 16 & \cve{CVE-2009-3550} & 6 & 8 & 42.86\\\hline
1st & 17 & \cve{CVE-2009-3243} & 13 & 23 & 36.11\\\hline
1st & 18 & \cve{CVE-2009-4377} & 12 & 22 & 35.29\\\hline
1st & 19 & \cve{CVE-2010-2285} & 6 & 11 & 35.29\\\hline
1st & 20 & \cve{CVE-2009-2559} & 6 & 11 & 35.29\\\hline
1st & 21 & \cve{CVE-2010-2284} & 6 & 12 & 33.33\\\hline
1st & 22 & \cve{CVE-2009-3242} & 7 & 16 & 30.43\\\hline
2nd & 1 & \cve{CVE-2009-3829} & 6 & 0 & 100.00\\\hline
2nd & 2 & \cve{CVE-2009-4376} & 6 & 0 & 100.00\\\hline
2nd & 3 & \cve{CVE-2010-0304} & 6 & 0 & 100.00\\\hline
2nd & 4 & \cve{CVE-2010-2286} & 6 & 0 & 100.00\\\hline
2nd & 5 & \cve{CVE-2010-2283} & 6 & 0 & 100.00\\\hline
2nd & 6 & \cve{CVE-2009-3551} & 6 & 0 & 100.00\\\hline
2nd & 7 & \cve{CVE-2009-3549} & 6 & 0 & 100.00\\\hline
2nd & 8 & \cve{CVE-2009-3241} & 16 & 8 & 66.67\\\hline
2nd & 9 & \cve{CVE-2009-2560} & 10 & 5 & 66.67\\\hline
2nd & 10 & \cve{CVE-2010-1455} & 44 & 10 & 81.48\\\hline
2nd & 11 & \cve{CVE-2009-2563} & 6 & 5 & 54.55\\\hline
2nd & 12 & \cve{CVE-2009-2562} & 6 & 5 & 54.55\\\hline
2nd & 13 & \cve{CVE-2009-2561} & 6 & 7 & 46.15\\\hline
2nd & 14 & \cve{CVE-2009-4378} & 6 & 7 & 46.15\\\hline
2nd & 15 & \cve{CVE-2010-2287} & 13 & 0 & 100.00\\\hline
2nd & 16 & \cve{CVE-2009-3550} & 6 & 8 & 42.86\\\hline
2nd & 17 & \cve{CVE-2009-3243} & 13 & 23 & 36.11\\\hline
2nd & 18 & \cve{CVE-2009-4377} & 12 & 22 & 35.29\\\hline
2nd & 19 & \cve{CVE-2010-2285} & 6 & 11 & 35.29\\\hline
2nd & 20 & \cve{CVE-2009-2559} & 6 & 11 & 35.29\\\hline
2nd & 21 & \cve{CVE-2010-2284} & 6 & 12 & 33.33\\\hline
2nd & 22 & \cve{CVE-2009-3242} & 8 & 15 & 34.78\\\hline
\end{tabular}

%% file: results/sate4/wireshark1-2-0-SATE-IV-5-train-test-run-quick-wireshark-cve-sdwt-spectrogram-graph-flucid.tex
\begin{tabular}{|c|r|l|c|c|r|}\hline
guess & run & algorithms & good & bad &  \% \\\hline
1st & 1 & \option{-nopreprep} \option{-sdwt} \option{-fft} \option{-diff} \option{-spectrogram} \option{-graph} \option{-flucid}  & 37 & 4 & 90.24\\\hline
1st & 2 & \option{-nopreprep} \option{-sdwt} \option{-fft} \option{-cheb} \option{-spectrogram} \option{-graph} \option{-flucid}  & 37 & 4 & 90.24\\\hline
1st & 3 & \option{-nopreprep} \option{-sdwt} \option{-fft} \option{-eucl} \option{-spectrogram} \option{-graph} \option{-flucid}  & 27 & 14 & 65.85\\\hline
1st & 4 & \option{-nopreprep} \option{-sdwt} \option{-fft} \option{-hamming} \option{-spectrogram} \option{-graph} \option{-flucid}  & 26 & 15 & 63.41\\\hline
1st & 5 & \option{-nopreprep} \option{-sdwt} \option{-fft} \option{-mink} \option{-spectrogram} \option{-graph} \option{-flucid}  & 22 & 19 & 53.66\\\hline
1st & 6 & \option{-nopreprep} \option{-sdwt} \option{-fft} \option{-cos} \option{-spectrogram} \option{-graph} \option{-flucid}  & 38 & 65 & 36.89\\\hline
2nd & 1 & \option{-nopreprep} \option{-sdwt} \option{-fft} \option{-diff} \option{-spectrogram} \option{-graph} \option{-flucid}  & 39 & 2 & 95.12\\\hline
2nd & 2 & \option{-nopreprep} \option{-sdwt} \option{-fft} \option{-cheb} \option{-spectrogram} \option{-graph} \option{-flucid}  & 39 & 2 & 95.12\\\hline
2nd & 3 & \option{-nopreprep} \option{-sdwt} \option{-fft} \option{-eucl} \option{-spectrogram} \option{-graph} \option{-flucid}  & 35 & 6 & 85.37\\\hline
2nd & 4 & \option{-nopreprep} \option{-sdwt} \option{-fft} \option{-hamming} \option{-spectrogram} \option{-graph} \option{-flucid}  & 29 & 12 & 70.73\\\hline
2nd & 5 & \option{-nopreprep} \option{-sdwt} \option{-fft} \option{-mink} \option{-spectrogram} \option{-graph} \option{-flucid}  & 31 & 10 & 75.61\\\hline
2nd & 6 & \option{-nopreprep} \option{-sdwt} \option{-fft} \option{-cos} \option{-spectrogram} \option{-graph} \option{-flucid}  & 39 & 64 & 37.86\\\hline
guess & run & class & good & bad &  \% \\\hline
1st & 1 & \cve{CVE-2009-3829} & 6 & 0 & 100.00\\\hline
1st & 2 & \cve{CVE-2009-2562} & 6 & 0 & 100.00\\\hline
1st & 3 & \cve{CVE-2009-4378} & 6 & 0 & 100.00\\\hline
1st & 4 & \cve{CVE-2010-2286} & 6 & 0 & 100.00\\\hline
1st & 5 & \cve{CVE-2010-0304} & 6 & 0 & 100.00\\\hline
1st & 6 & \cve{CVE-2009-4376} & 6 & 0 & 100.00\\\hline
1st & 7 & \cve{CVE-2010-2283} & 6 & 0 & 100.00\\\hline
1st & 8 & \cve{CVE-2009-3551} & 6 & 0 & 100.00\\\hline
1st & 9 & \cve{CVE-2009-3550} & 6 & 0 & 100.00\\\hline
1st & 10 & \cve{CVE-2009-3549} & 6 & 0 & 100.00\\\hline
1st & 11 & \cve{CVE-2009-2563} & 6 & 2 & 75.00\\\hline
1st & 12 & \cve{CVE-2009-2560} & 11 & 4 & 73.33\\\hline
1st & 13 & \cve{CVE-2009-3241} & 15 & 9 & 62.50\\\hline
1st & 14 & \cve{CVE-2010-1455} & 31 & 23 & 57.41\\\hline
1st & 15 & \cve{CVE-2009-2561} & 6 & 6 & 50.00\\\hline
1st & 16 & \cve{CVE-2010-2287} & 6 & 6 & 50.00\\\hline
1st & 17 & \cve{CVE-2009-2559} & 6 & 6 & 50.00\\\hline
1st & 18 & \cve{CVE-2009-3243} & 16 & 16 & 50.00\\\hline
1st & 19 & \cve{CVE-2010-2285} & 6 & 7 & 46.15\\\hline
1st & 20 & \cve{CVE-2009-4377} & 12 & 16 & 42.86\\\hline
1st & 21 & \cve{CVE-2010-2284} & 6 & 9 & 40.00\\\hline
1st & 22 & \cve{CVE-2009-3242} & 6 & 17 & 26.09\\\hline
2nd & 1 & \cve{CVE-2009-3829} & 6 & 0 & 100.00\\\hline
2nd & 2 & \cve{CVE-2009-2562} & 6 & 0 & 100.00\\\hline
2nd & 3 & \cve{CVE-2009-4378} & 6 & 0 & 100.00\\\hline
2nd & 4 & \cve{CVE-2010-2286} & 6 & 0 & 100.00\\\hline
2nd & 5 & \cve{CVE-2010-0304} & 6 & 0 & 100.00\\\hline
2nd & 6 & \cve{CVE-2009-4376} & 6 & 0 & 100.00\\\hline
2nd & 7 & \cve{CVE-2010-2283} & 6 & 0 & 100.00\\\hline
2nd & 8 & \cve{CVE-2009-3551} & 6 & 0 & 100.00\\\hline
2nd & 9 & \cve{CVE-2009-3550} & 6 & 0 & 100.00\\\hline
2nd & 10 & \cve{CVE-2009-3549} & 6 & 0 & 100.00\\\hline
2nd & 11 & \cve{CVE-2009-2563} & 6 & 2 & 75.00\\\hline
2nd & 12 & \cve{CVE-2009-2560} & 12 & 3 & 80.00\\\hline
2nd & 13 & \cve{CVE-2009-3241} & 16 & 8 & 66.67\\\hline
2nd & 14 & \cve{CVE-2010-1455} & 43 & 11 & 79.63\\\hline
2nd & 15 & \cve{CVE-2009-2561} & 6 & 6 & 50.00\\\hline
2nd & 16 & \cve{CVE-2010-2287} & 12 & 0 & 100.00\\\hline
2nd & 17 & \cve{CVE-2009-2559} & 6 & 6 & 50.00\\\hline
2nd & 18 & \cve{CVE-2009-3243} & 19 & 13 & 59.38\\\hline
2nd & 19 & \cve{CVE-2010-2285} & 6 & 7 & 46.15\\\hline
2nd & 20 & \cve{CVE-2009-4377} & 12 & 16 & 42.86\\\hline
2nd & 21 & \cve{CVE-2010-2284} & 6 & 9 & 40.00\\\hline
2nd & 22 & \cve{CVE-2009-3242} & 8 & 15 & 34.78\\\hline
\end{tabular}

%% file: results/sate4/wireshark1-2-0-SATE-IV-5-train-test-run-quick-wireshark-cwe-sdwt-flucid.tex
\begin{tabular}{|c|r|l|c|c|r|}\hline
guess & run & algorithms & good & bad &  \% \\\hline
1st & 1 & \option{-cweid} \option{-nopreprep} \option{-sdwt} \option{-fft} \option{-diff} \option{-flucid}  & 31 & 5 & 86.11\\\hline
1st & 2 & \option{-cweid} \option{-nopreprep} \option{-sdwt} \option{-fft} \option{-eucl} \option{-flucid}  & 29 & 7 & 80.56\\\hline
1st & 3 & \option{-cweid} \option{-nopreprep} \option{-sdwt} \option{-fft} \option{-mink} \option{-flucid}  & 17 & 19 & 47.22\\\hline
1st & 4 & \option{-cweid} \option{-nopreprep} \option{-sdwt} \option{-fft} \option{-hamming} \option{-flucid}  & 14 & 22 & 38.89\\\hline
2nd & 1 & \option{-cweid} \option{-nopreprep} \option{-sdwt} \option{-fft} \option{-diff} \option{-flucid}  & 33 & 3 & 91.67\\\hline
2nd & 2 & \option{-cweid} \option{-nopreprep} \option{-sdwt} \option{-fft} \option{-eucl} \option{-flucid}  & 34 & 2 & 94.44\\\hline
2nd & 3 & \option{-cweid} \option{-nopreprep} \option{-sdwt} \option{-fft} \option{-mink} \option{-flucid}  & 27 & 9 & 75.00\\\hline
2nd & 4 & \option{-cweid} \option{-nopreprep} \option{-sdwt} \option{-fft} \option{-hamming} \option{-flucid}  & 23 & 13 & 63.89\\\hline
guess & run & class & good & bad &  \% \\\hline
1st & 1 & CWE399 & 4 & 0 & 100.00\\\hline
1st & 2 & CWE189 & 4 & 0 & 100.00\\\hline
1st & 3 & \cwe{NVD-CWE-Other} & 11 & 1 & 91.67\\\hline
1st & 4 & CWE20 & 30 & 10 & 75.00\\\hline
1st & 5 & \cwe{NVD-CWE-noinfo} & 34 & 34 & 50.00\\\hline
1st & 6 & CWE119 & 8 & 8 & 50.00\\\hline
2nd & 1 & CWE399 & 4 & 0 & 100.00\\\hline
2nd & 2 & CWE189 & 4 & 0 & 100.00\\\hline
2nd & 3 & \cwe{NVD-CWE-Other} & 11 & 1 & 91.67\\\hline
2nd & 4 & CWE20 & 34 & 6 & 85.00\\\hline
2nd & 5 & \cwe{NVD-CWE-noinfo} & 53 & 15 & 77.94\\\hline
2nd & 6 & CWE119 & 11 & 5 & 68.75\\\hline
\end{tabular}

%% file: results/sate4/wireshark1-2-0-SATE-IV-5-train-test-run-quick-wireshark-cwe-low-flucid.tex
\begin{tabular}{|c|r|l|c|c|r|}\hline
guess & run & algorithms & good & bad &  \% \\\hline
1st & 1 & \option{-cweid} \option{-nopreprep} \option{-low} \option{-fft} \option{-diff} \option{-flucid}  & 30 & 6 & 83.33\\\hline
1st & 2 & \option{-cweid} \option{-nopreprep} \option{-low} \option{-fft} \option{-cheb} \option{-flucid}  & 30 & 6 & 83.33\\\hline
1st & 3 & \option{-cweid} \option{-nopreprep} \option{-low} \option{-fft} \option{-eucl} \option{-flucid}  & 25 & 11 & 69.44\\\hline
1st & 4 & \option{-cweid} \option{-nopreprep} \option{-low} \option{-fft} \option{-mink} \option{-flucid}  & 20 & 16 & 55.56\\\hline
1st & 5 & \option{-cweid} \option{-nopreprep} \option{-low} \option{-fft} \option{-cos} \option{-flucid}  & 36 & 40 & 47.37\\\hline
1st & 6 & \option{-cweid} \option{-nopreprep} \option{-low} \option{-fft} \option{-hamming} \option{-flucid}  & 12 & 24 & 33.33\\\hline
2nd & 1 & \option{-cweid} \option{-nopreprep} \option{-low} \option{-fft} \option{-diff} \option{-flucid}  & 31 & 5 & 86.11\\\hline
2nd & 2 & \option{-cweid} \option{-nopreprep} \option{-low} \option{-fft} \option{-cheb} \option{-flucid}  & 31 & 5 & 86.11\\\hline
2nd & 3 & \option{-cweid} \option{-nopreprep} \option{-low} \option{-fft} \option{-eucl} \option{-flucid}  & 30 & 6 & 83.33\\\hline
2nd & 4 & \option{-cweid} \option{-nopreprep} \option{-low} \option{-fft} \option{-mink} \option{-flucid}  & 22 & 14 & 61.11\\\hline
2nd & 5 & \option{-cweid} \option{-nopreprep} \option{-low} \option{-fft} \option{-cos} \option{-flucid}  & 48 & 28 & 63.16\\\hline
2nd & 6 & \option{-cweid} \option{-nopreprep} \option{-low} \option{-fft} \option{-hamming} \option{-flucid}  & 16 & 20 & 44.44\\\hline
guess & run & class & good & bad &  \% \\\hline
1st & 1 & CWE399 & 6 & 1 & 85.71\\\hline
1st & 2 & CWE20 & 48 & 12 & 80.00\\\hline
1st & 3 & \cwe{NVD-CWE-Other} & 18 & 7 & 72.00\\\hline
1st & 4 & CWE189 & 6 & 3 & 66.67\\\hline
1st & 5 & \cwe{NVD-CWE-noinfo} & 61 & 61 & 50.00\\\hline
1st & 6 & CWE119 & 14 & 19 & 42.42\\\hline
2nd & 1 & CWE399 & 6 & 1 & 85.71\\\hline
2nd & 2 & CWE20 & 48 & 12 & 80.00\\\hline
2nd & 3 & \cwe{NVD-CWE-Other} & 18 & 7 & 72.00\\\hline
2nd & 4 & CWE189 & 6 & 3 & 66.67\\\hline
2nd & 5 & \cwe{NVD-CWE-noinfo} & 78 & 44 & 63.93\\\hline
2nd & 6 & CWE119 & 22 & 11 & 66.67\\\hline
\end{tabular}